\documentclass{aa}
\usepackage{graphicx}
\usepackage{natbib}
\usepackage{amssymb}
\usepackage{amstext}
\usepackage{latexsym} 
\usepackage[sumlimits]{amsmath}
\newcommand{\Ha}{H$\alpha$}			
\newcommand{\NII}{[N{\sc ii}]}			
\newcommand{\HII}{H{\sc ii}}			
\newcommand{\HI}{H{\sc i}}			

\begin{document}
   \title{The H$\alpha$ Galaxy Survey \thanks{
Based on observations made with the Jacobus Kapteyn Telescope operated 
on the island of La Palma by the Isaac Newton Group in the Spanish 
Observatorio del Roque de los Muchachos of the Instituto de Astrof\'\i sica 
de Canarias
   }}
   \subtitle{VI. Star-forming companions of nearby field galaxies}
   \author{P.~A. James 
   \inst{1},
    J. O'Neill
    \inst{2},
     N.~S. Shane
    \inst{1,3},
    } \offprints{P.A. James} 
          \institute{Astrophysics Research
	  Institute, Liverpool John Moores University, Twelve Quays
	  House, Egerton Wharf, Birkenhead CH41 1LD, UK \\
	  \email{paj@astro.livjm.ac.uk} 
	  \and Wirral Grammar School for Girls, Heath Road, 
          Bebington, Wirral CH63 3AF, UK\\
	  \and Planetary Science Group, Mullard Space Science Laboratory,
          Holmbury St. Mary, Dorking, Surrey RH5 6NT, UK\\
          }
          \date{Received            ; accepted               }

\abstract 
{} 
{We searched for star-forming satellite galaxies that are
close enough to their parent galaxies to be considered analogues
of the Magellanic Clouds.}
{Our search technique relied on the detection of the satellites in
continuum-subtracted narrow-band \Ha\ imaging of the central galaxies,
which removes most of the background and foreground line-of-sight
companions, thus giving a high probability that we are detecting true
satellites.  The search was performed for 119 central galaxies at
distances between 20 and 40~Mpc, although spatial incompleteness means
that we have effectively searched 53 full satellite-containing
volumes.}
{We find only 9 `probable' star-forming satellites, around 9 different 
central galaxies, and 2 more `possible' satellites.  
After incompleteness correction, this is equivalent to 
0.17/0.21  satellites per central galaxy.  
This frequency is unchanged whether we
consider all central galaxy types or just those of Hubble types S0a -
Sc, i.e.  only the more luminous and massive spiral types.  The
satellites found are generally similar to the Magellanic Clouds and to
field Sm and Im galaxies, in terms of their normalised star formation
rates.  However, this conclusion is somewhat circular as the similarity of 
properties to known Sm/Im galaxies was used as a classification criterion.
The Small Magellanic Cloud is just below the median values of
both star formation rate and $R$-band luminosity of the 9 probable
satellites. The Large Magellanic Cloud, however, has a higher $R$-band
luminosity than any of the 9 and is only exceeded in star formation
rate by the one satellite that appears to be undergoing a
tidally-induced starburst.  Thus the Milky Way appears to be quite
unusual, both in having two star-forming satellite galaxies and in
the high luminosity of the Large Magellanic Cloud.} 
{}


\authorrunning{James et al.}
\titlerunning{H$\alpha$ Galaxy Survey. VI.}
\maketitle
%
\section{Introduction}

Satellite galaxies are of great importance in our understanding of the
formation and morphological evolution of all types of galaxy.  Minor
mergers of disk and dwarf galaxies provide one possible process for
forming or enlarging bulges \citep[e.g.][]{hayn00,kann04,elic06}, and
the associated tidal perturbation is likely to distort and ultimately
thicken disks \citep{quin86,quin93}. Indeed, several minor mergers may
suffice to convert a disk galaxy into an elliptical
\citep{bour07,kavi07}. Gas-rich dwarfs may provide gas reservoirs for
rejuvenating early-type disk galaxies and thus prolonging or
re-establishing star formation (SF) activity \citep{whit91,hau07},
whilst also changing the metallicity of the remaining gas component.
Finally, accretion of dwarf galaxies may be very significant for the
formation of the stellar haloes of galaxies
\citep[e.g.][]{sear78,read06}.

However, in order to quantify the typical effects of such mergers on
disk galaxies, it is necessary to know what fraction actually have
close companions, and how many of these contain significant gas
reservoirs.  These fractions, when combined with a typical time
for the decay of orbits through dynamical friction, should ultimately 
enable an
estimate to be made of the minor merger rate, and the resulting impact
on the SF of the central galaxies.  

A starting-point for
such an investigation is provided by the Milky Way, which has a
significant number of satellite galaxies.  Of these, most are very
faint and low-surface brightness dwarf spheroidal and dwarf irregular
galaxies; in terms of mass, and future evolutionary impact, by far the
most dominant of the probable satellites are the Large and Small Magellanic
Clouds (LMC and SMC henceforth).  Since the Milky Way is often assumed
to be a typical field galaxy, probably of type SBb - SBc, with a total
luminosity close to the characteristic value L$^{\star}$ in the
\cite{sche76} luminosity function, it might be natural to assume that
Magellanic Cloud-like companions are also typical of the field galaxy
population.  Indeed, there are several well-studied bright satellites
around local galaxies, such as M~32 and NGC~205, the early-type
companions of M~31.  However, no systematic search for Magellanic-type
companions around a representative sample of field galaxies has yet
been carried out, so it is not clear whether the Milky Way is typical
or unusual in having two near neighbours of this type.

Some studies of field galaxy companions have been presented in the
literature.  \citet{zari97} searched for companions at least 2.2~mag
fainter than the primary galaxy, within 1000~km~s$^{-1}$ in velocity
and 500~kpc in projected separation, and found 115 satellites around
69 primary galaxies.  However, the allowed separations are very much
greater than the separation of the Magellanic Clouds from the Milky
Way, and \citet{zari97} comment that some of the primaries have four
or five companions, and these systems may thus be better regarded as
galaxy groups. Only about 10\% of the companions have projected
separations less than 70~kpc, potentially comparable to the
Magellanic Clouds.  \citet{noes01} found $\sim$30\% of a sample of
field dwarf galaxies to have companions within a projected separation
of 100~kpc and a recession velocity difference of
$\pm500$~km~s$^{-1}$.  \citet{mado04}, following a pioneering study by
\citet{both77}, performed a search for companions around isolated
elliptical galaxies.  Within a 75~kpc projected radius they found
1.0$\pm$0.5 companions per elliptical galaxy, higher than the
0.12$\pm$0.42 companions per galaxy found by \citet{both77}.

One requirement of any such study is for recession velocity
information to remove line-of-sight pairs, which generally dominate
the population of apparent pairs found in pure imaging studies,
particularly for blue companions \citep{chen07}.  A complicating
factor in obtaining such data through spectroscopic surveys is the
problem of fibre collisions with multi-object spectrographs, which
means that close pairs are under-represented in many existing surveys.
Here we undertake a search using an alternative method based on
narrow-band \Ha\ imaging of the areas around field galaxies in the
local Universe.  Specifically, we search for star forming companions
of these galaxies, which will show up via \Ha\ line emission, and as a
result this study says nothing about non-star forming companions.  Any
companions detected in \Ha\ are likely, given the $\sim$50 Angstrom
width of \Ha\ filters used, to be truly associated with central target
galaxy.  The velocity range for \Ha\ to lie within the same
narrow-band filter is $\pm$1000~km~s$^{-1}$, which excludes projected
companions with good efficiency, but the number of detected companions
will be an upper limit as there will still be some line-of-sight
projections within this range. Thus the radial distance range in which
satellites could lie is 29 Mpc in depth for an assumed Hubble constant
of 70~km~s$^{-1}$Mpc$^{-1}$; as explained in section
\ref{sec:galstats}, we search for satellites to a projected distance
of 75kpc in the plane of the sky, giving a total search volume of
$\sim$0.5~Mpc$^{3}$. Of course, any `true' satellites will lie in a
much smaller volume than this about the central galaxy.  There is also
the possibility of including very distant background galaxies, with
other emission lines redshifted into the \Ha\ filter bandpass.

Whilst this technique will clearly detect only the line-emitting
fraction of all satellite galaxies, it is important to note recent
results demonstrating that this is equivalent to the entire gas-rich
population in the equivalent mass range, i.e. that there are no or
very few quiescent gas-rich galaxies.  \cite{meur06} found that all 93
of the \HI -selected galaxies in their sample were detected in \Ha\
emission. \citet{hain07}, using SDSS data, found all of the $\sim600$
low-luminosity galaxies studied ($-18 < M_r < -16$) in the
lowest-density environments to have \Ha\ emission.  The previous paper
in this series \citep{paper5} found that all
117 late-type Sm and Im galaxies in the \Ha GS sample are actively
forming stars, and these correspond to more than 50\% of the galaxies
of these types satisfying the selection criteria of this sample. Thus
possession of significant gas reservoirs and star formation (with
corresponding \Ha\ emission) appear to be synonymous.

\section{Data and methods}
\label{sec:method}

We use imaging data from the \Ha\ Galaxy Survey \citep{paper1}, \Ha GS
henceforth. This survey contains data for 327 galaxies selected from
the Uppsala General Catalogue of Galaxies \citep{nils73} (UGC) to have
diameters between 1\farcm7 and 6\farcm0, measured recession velocities
of less than 3000~km~s$^{-1}$, major-to-minor axis ratios less than
4.0, and Hubble type later than S0a.  The Virgo cluster core was
excluded, so this is effectively a field galaxy sample, but otherwise
it should be unbiased with respect to presence or absence of
companion or satellite galaxies. 

The \Ha GS data comprise broad-band $R$ imaging, and \Ha\ in one of
several narrow band filters, selected to match the recession velocity
of the target UGC galaxy, obtained at the 1.0~metre Jacobus Kapteyn
Telescope.  The field of view of the JKT CCD camera was $\sim$
11$\times$11~arcmin, but we conservatively use only the central
9.5$\times$9.5~arcmin area to avoid vignetting and cosmetic problems
with the edges of the frames.  The \Ha\ images used here were
continuum-subtracted, using either observations in an
intermediate-width continuum filter, or scaled $R$-band exposures, and
flux calibrated, as outlined in \citet{paper1}.  All galaxy distances
(from Virgo-infall corrected recession velocities) and $R$-band
magnitudes for the central galaxies used in the present analysis are
those listed in Table 3 of \citet{paper1}.  However, for consistency
with the analysis presented in \citet{paper5}
(hereafter referred to as paper V), the SF rates
for central galaxies used here are multiplied by a factor 0.7,
consistent with the assumption of a `Salpeter light' stellar initial
mass function, rather than the Salpeter function assumed by
\cite{kenn98} and adopted in paper I.

All \Ha GS galaxies with distances greater than or equal to 20~Mpc
were included in this analysis; this was a total of 119 central
galaxies.  Galaxies within this distance limit were excluded because
of the small effective volume included within the CCD field of view.
The method employed when searching for potential satellite galaxies
was initially to scan the continuum-subtracted \Ha\ images, looking
for any apparent regions of emission that are detached from the main
galaxy.  Any potential sources found were then `blinked' with the
$R$-band image (which is aligned to sub-pixel accuracy) to check for,
e.g., poorly-subtracted foreground stars, cosmetic defects or bright
cosmic ray trails.  The brightest of such spurious sources can be
excluded from consideration if they are not present in the $R$-band
image (since this filter includes the \Ha\ emission), and they can
also be excluded if not at least as extended as the stellar point
spread function.  In most of the frames studied, the situation was
completely unambiguous, as no possible satellite galaxies were seen on
the \Ha\ frame, and almost all of the spurious objects were quickly
identified as such.  This left 32 cases where an apparently real
emission-line source was identified somewhere within the CCD field of
view.

One possible source of incompleteness concerns those galaxies for
which the \Ha\ line from the central galaxy lies close to the edge of
the bandpass of the narrow band filter used.  There is significant
overlap between the redshifted \Ha\ filters that were available for
use with the JKT, so this was not severe problem, but there is a small
effect for those galaxies lying close to the `changeover' recession
velocity from one filter to the next.  To quantify this, the filter
throughputs were calculated for each galaxy recession velocity, and
for velocities 300~km~s$^{-1}$ on either side of this to represent the
kinematic limits of likely satellite populations.  The most severe
bias found was for UGC~4260, where the galaxy itself lies just on the
`wrong' side of a changeover point; adding 300~km~s$^{-1}$ to this
velocity drops the filter throughput to 62\% of the value at the
galaxy recession velocity.  There are 4 cases where this throughput
drop at $\pm$300~km~s$^{-1}$ is 69 or 70\%, 19 cases from 71 to 75\%,
and 7 cases from 76 to 80\%.  For the remaining 88 out of 119
galaxies, the throughput across the entire likely satellite velocity
range is greater than 80\% of that at the galaxy recession velocity.

The flux limit for our \Ha\ images was calculated from faint sources
in real images, including photon noise and systematic effects due to
irregularities in the background `sky' regions.  A moderately extended
source like our putative companions would be detected at 5$\sigma$ for
an observed \Ha\ plus \NII\ flux of 4.3$\times 10^{-18}$~W~m$^{-2}$.
This corresponds to a SF rate of 0.004 -- 0.0076~M$_{\odot}$yr$^{-1}$
for a galaxy at a distance of 30~Mpc, with the exact value in the
range depending on the extinction corrections adopted.

The remaining, and distinctly problematic issue concerning the 32
putative satellites was to distinguish whether these sources were
truly separate galaxies, or just outlying SF regions of the central
galaxy.  Conservatively, all 32 sources were included for further
analysis, even where the appearance on the $R$-band image pointed
towards the latter being the case, and resolution of this question was
left to further analysis of the source properties in section
\ref{sec:satclass} below.  This analysis was based on total \Ha\ and
$R$-band fluxes of the companions, which were measured using matched
elliptical apertures.  The \Ha\ fluxes were then converted to SF rates
using the conversion formula of \citet{kenn98} scaled by the 0.7
factor mentioned above, but with internal extinction corrections based
on companion object $M_R$-magnitudes using the methods of
\citet{helm04}.

\section{Classification of putative satellite galaxies}
\label{sec:satclass}

\begin{figure}
\centering
\rotatebox{-90}{
\includegraphics[height=8.5cm]{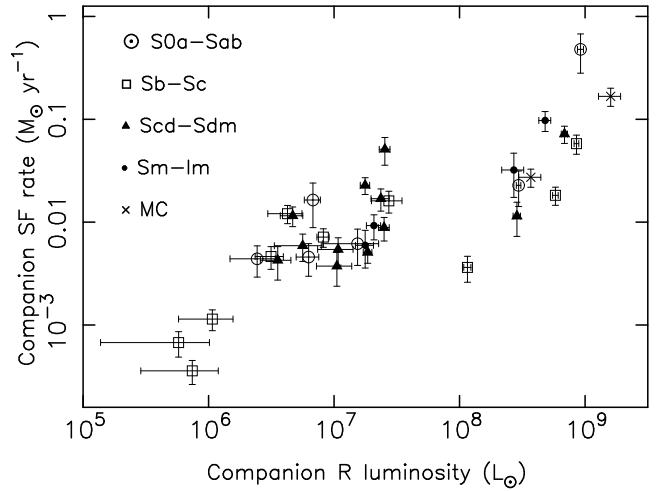}
}
\caption{
SF rate vs $R$-band luminosity for the 32 possible
companion galaxies, with the Magellanic Cloud properties shown for
comparison (crosses).  The different points show the Hubble type
of the central UGC galaxy.  
}
\label{fig:dsfvdrt}
\end{figure}

The first stage in separating the 32 \Ha -emitting sources into true
companions and outlying \HII\ regions was to examine their
distributions of SF rate and $R$-band luminosity, as shown in Fig.
\ref{fig:dsfvdrt}.  The range of SF rates is broadly consistent with
those found in Magellanic dwarf galaxies, but in the $R$-band
luminosity distribution there is already the hint of bimodality, with
two components around $10^7$ and $10^{8.5}~L_{\odot}$.  The Magellanic
Clouds are included in this plot (crosses), with the SF rates having been
calculated from the \Ha\ data of \citet{kenn86}, using the same
method of correcting for internal extinction as was applied to the SF
rates of the putative satellite galaxies.  Both Magellanic Clouds lie 
in the group of higher luminosity objects in Fig. \ref{fig:dsfvdrt}.

\begin{figure}
\centering
\rotatebox{-90}{
\includegraphics[height=8.5cm]{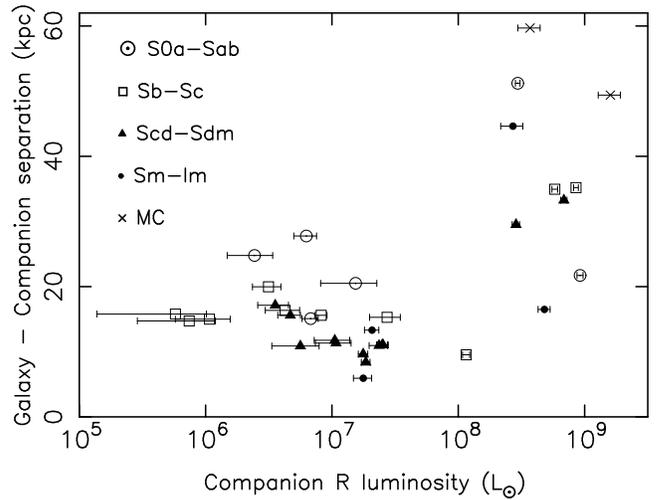}
}
\caption{
Separation of the 
companion object and central galaxy in kpc vs the $R$-band luminosity of
the companion.  The separation plotted is the actual
distance for the Magellanic Clouds, and the projected separation on the 
sky for the other objects.
}
\label{fig:dgdvdRT}
\end{figure}

Figure \ref{fig:dgdvdRT} shows the projected separation of the
putative companion from the nucleus of the central galaxy, plotted
against the companion $R$-band luminosity in solar units. This figure
adds weight to the supposition that the candidate companions with
lower luminosities in the $R$-band are \HII\ regions rather than
separate galaxies, as they show a strong tendency to lie closer to the
central galaxy than do the brighter sources.

A further plot used to discriminate between \HII\ regions and
satellite galaxies makes use of the SF timescale, a parameter explored
extensively in paper V.  The SF timescale is given by the total stellar
mass of the galaxy divided by the current SF rate, such that a
constant SF rate throughout a galaxy's history would result in a SF
timescale of a Hubble time.  As explained in paper V, the total
stellar mass is derived from $R$-band photometry, with an adopted
$R$-band mass-to-light ratio for late-type dwarfs of 0.65, based on
the models of \cite{bell01}.  The timescale is increased by a factor
of 1.67 to account for gas recycling of 40\% of the mass of any new
generation of stars \citep{vanz01}.

\begin{figure}
\centering
\rotatebox{-90}{
\includegraphics[height=8.5cm]{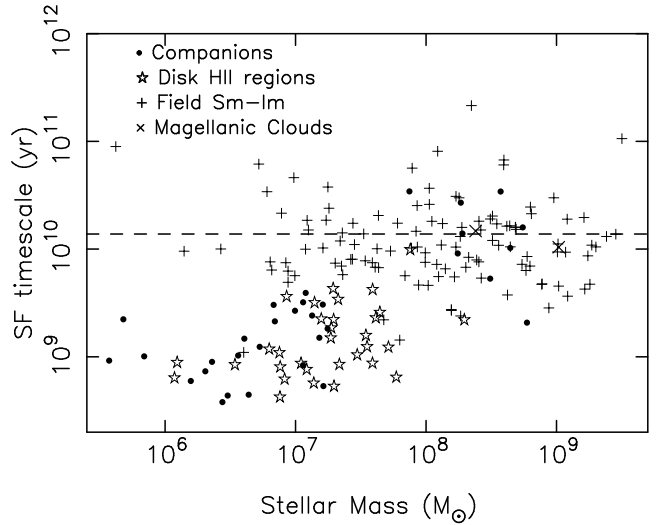}
}
\caption{
The time required to form the total stellar mass of the 
system, 
for the putative satellite galaxies, the Magellanic Clouds,  
field Sm and Im galaxies from the \Ha GS survey, and disk \HII\ regions,
vs stellar mass.
}
\label{fig:sftsat}
\end{figure}

As was concluded in paper V, Fig. \ref{fig:sftsat} shows the field Sm
and Im late-type galaxies from the \Ha GS sample to scatter around a
mean SF timescale of just under a Hubble time, indicating
approximately constant SF activity in these galaxies.  The crosses
show the positions of the LMC and SMC within
this distribution, where the Magellanic Cloud values are based on SF
rates from the \Ha\ measurements of \cite{kenn86}.  Interestingly, both
lie close to the centre of the range of SF timescales of the field
galaxies, with values of $\sim$10--15 Gyr.

The brightest, and hence highest mass, of the putative satellite
galaxies also lie within the distribution of field galaxies shown in
Fig. \ref{fig:sftsat}, and thus appear similar in their stellar masses
and current SF activity to field Sm and Im galaxies.  The one bright
satellite which is displaced somewhat to a short SF timescale is the
companion to UGC~4541, which appears tidally disturbed and may be
undergoing an interaction-induced starburst or other nuclear activity.
However, {\em all} of the putative companions with stellar masses
below $\sim 5 \times 10^7~M_{\odot}$ have short SF timescales of
$<$10$^{10}$~yr, and the average for these objects is $<$10$^9$~yr.  This
is most simply explained if these objects are SF regions in the outer
disks of the central \Ha GS galaxies, which would naturally have low
SF timescales since they represent short-lived local enhancements in
the SF rate of their host galaxies. This is confirmed by the position
of known \HII\ regions from the disk regions of \Ha GS galaxies, shown as
stars in Fig. \ref{fig:sftsat}. The existence of SF at a low
level in the outer regions of disk galaxies has also been noted in 
UV imaging from the GALEX mission \citep{thil05,gild05}.

\begin{figure}
\centering
\rotatebox{-90}{
\includegraphics[height=8.5cm]{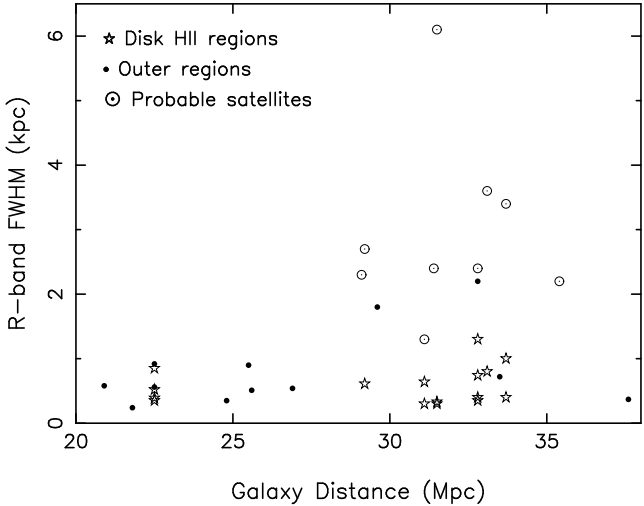}
}
\caption{
The FWHM sizes in kpc of the 9 satellite galaxies, a selection of disk
\HII\ regions, and the outer disk regions, measured from $R$-band images.
}
\label{fig:Rsatsize}
\end{figure}

\begin{figure}
\centering
\rotatebox{-90}{
\includegraphics[height=8.5cm]{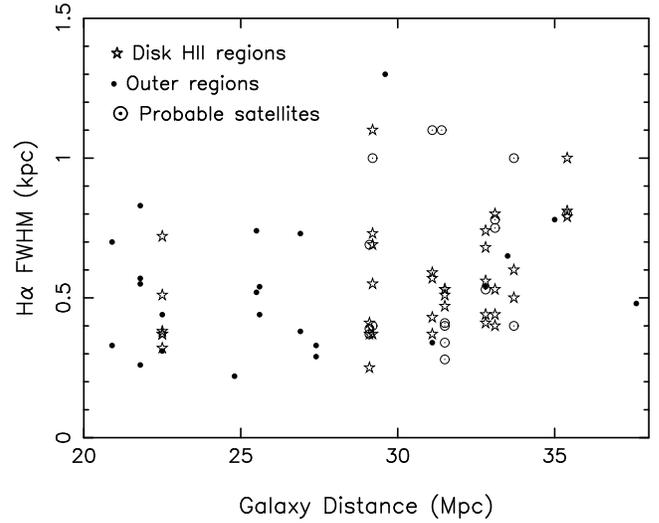}
}
\caption{
The FWHM sizes in kpc of the \Ha\ emission-line regions detected by SExtractor
within the 9 satellite galaxies and the outer disk regions, compared with the
equivalent dimension for disk \HII\ regions.
}
\label{fig:Hasatsize}
\end{figure}

However, even though Figs. \ref{fig:dsfvdrt} and \ref{fig:sftsat}
indicate that the 32 objects split into two groups as described in the
previous paragraph, they are not completely conclusive.  In
particular, objects in the high luminosity/mass end of the
distribution of probable outer \HII\ regions lie on the outskirts of
the distribution of field Sm and Im galaxies in Fig. \ref{fig:sftsat}.
As a final test, the sizes of the objects were measured from our
$R$-band and \Ha\ images, and plotted in Figs. \ref{fig:Rsatsize} and
\ref{fig:Hasatsize}.  In both figures, the sizes were determined in a
completely automated and objective fashion, using Full Width at Half
Maximum (FWHM) values from the SExtractor package.  The resulting
$R$-band sizes are plotted against the distance to the central galaxy
in Mpc, in Fig. \ref{fig:Rsatsize}.  This confirms that all 9 of the
likely satellite galaxies are very extended in terms of their
continuum emission, with FWHM sizes of 1.5 -- 6.0~kpc.  Many of the
probable outer disk sources are too faint in the $R$-band to be picked
up as separate sources by the SExtractor software; most of the
remainder have $R$-band FWHM sizes between 0.2 and 1.0~kpc, identical
to the range of sizes of disk \HII\ regions, also shown in
Fig. \ref{fig:Rsatsize} for comparison.  However, two of the `outer
disk' objects do show significant extents in the $R$-band images.
These are objects associated with UGC~3530 and UGC~4260, with $R$-band
extents of 1.8 and 2.2~kpc respectively.  Both objects are amongst the
most massive of the 23 `outer disk' objects, at just over
10$^7$~M$_{\odot}$, and have star-formation timescales of $\sim$1 and
2.5~Gyr respectively.  Thus these may be satellite galaxies in a
star-bursting phase, and we will keep them in our analysis as
`possible' satellites.  Images of these objects are shown in Appendix
A, with the putative satellites indicated by lines above and to one
side of their location.
  
The \Ha\ sizes of objects, plotted on the y-axis in Fig. \ref{fig:Hasatsize},
show no significant dependence on object type.  The SExtractor algorithm split
the likely satellite galaxies into separate resolved \HII\ regions in several
cases, and the resulting regions had FWHM sizes consistent with those of both
the 23 outlying regions (all of which were detected in the \Ha\ analysis), and 
a selection of disk \HII\ regions.  The mean FWHM of 36 disk \HII\ regions
was 0.55~kpc (standard deviation 0.19~kpc, standard error on the mean 
0.03~kpc); for the 23 outlying regions, the mean was 0.55~kpc (s.d. 0.24~kpc,
s.e. 0.05~kpc); and for the 9 probable satellites, which were detected as
17 \HII\ regions, the mean FWHM was 0.63~kpc (s.d. 0.29~kpc, s.e. 0.07~kpc).

Henceforth, we
consider only the 9 companion objects with $R$-band luminosities
greater than $10^8$~L$_{\odot}$ to be `probable' satellite galaxies, and
the companions to UGC~3530 and UGC~4260 as `possible' satellites.  The 
remaining 21 objects are consistent with being outer disk \HII\ regions
in terms of all the parameters considered here.  However, they will be 
targetted for spectroscopic follow-up in future work (outlined in 
section \ref{sec:disc} of this paper).

\begin{table*}
\begin{center} 
\begin{tabular}{rlccccccccccc}
\hline
\hline
UGC$_{\rm c}$ &  Type$_{\rm c}$  & Dist &  SFR$_{\rm c}$ &  $\delta$SFR$_{\rm c}$  &  $R_{\rm tot,c}$ 
&  $\delta R_{\rm tot,c}$ &  L$_{R,{\rm c}}$ &  
SFR$_{\rm s}$ & $\delta$SFR$_{\rm s}$  &  L$_{R, {\rm s}}$ &  $\delta$L$_{R, {\rm s}}$  &   Sepn \cr
  &  & Mpc & M$_{\odot}$~yr$^{-1}$ & & mag &  & L$_{\odot}$ & M$_{\odot}$~yr$^{-1}$ &  & L$_{\odot}$ 
&  & kpc \cr
\hline
 2603 & Im   & 33.7 &  0.28 & 0.10 & 14.59 & 0.06 & 9.70(08) & 0.032 & 0.015 & 2.71(8) & 5.4(7) & 44.6\\
 4273 & SBb  & 35.4 &  1.92 & 0.36 & 11.84 & 0.04 & 1.36(10) & 0.018 & 0.004 & 5.81(8) & 3.0(7) & 35.0\\
 4362 & S0a  & 33.1 &  0.46 & 0.17 & 11.96 & 0.04 & 1.06(10) & 0.023 & 0.009 & 2.96(8) & 1.2(7) & 51.2\\
 4469 & SBcd & 31.5 &  1.52 & 0.25 & 12.50 & 0.04 & 5.81(09) & 0.072 & 0.014 & 6.86(8) & 3.9(7) & 33.3\\
 4541 & Sa   & 31.4 &  0.21 & 0.06 & 11.37 & 0.04 & 1.63(10) & 0.478 & 0.197 & 9.21(8) & 5.0(7) & 21.7\\
 4574 & SBb  & 31.1 &  2.96 & 0.60 & 11.26 & 0.04 & 1.77(10) & 0.058 & 0.012 & 8.55(8) & 3.7(7) & 35.2\\
 5688 & SBm  & 29.2 &  0.44 & 0.08 & 13.57 & 0.05 & 1.87(09) & 0.097 & 0.021 & 4.81(8) & 5.2(7) & 16.5\\
 6506 & SBd  & 29.1 &  0.06 & 0.02 & 14.74 & 0.06 & 6.31(08) & 0.011 & 0.004 & 2.86(8) & 1.9(7) & 29.5\\
12788 & Sc   & 32.8 &  1.54 & 0.30 & 12.51 & 0.04 & 6.25(09) & 0.004 & 0.001 & 1.16(8) & 7.6(6) &  9.5\\
\hline
\end{tabular}
\caption[]{Main properties of the central and satellite galaxies}
\label{tbl:galprop}
\end{center}
\end{table*}

\begin{figure}
\centering
\rotatebox{-90}{
\includegraphics[height=8.5cm]{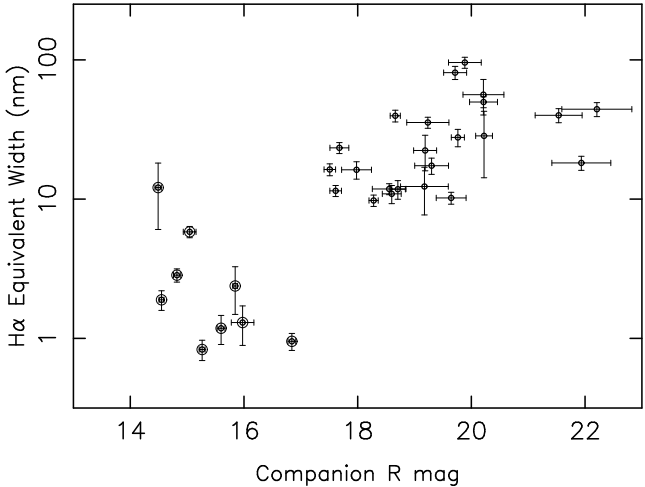}
}
\caption{
The \Ha\ Equivalent Width values for the 9 satellite galaxies and 
the outer disk regions, plotted against total $R$-band mag.
}
\label{fig:ewvR}
\end{figure}

\section{Properties of the 9 probable satellite galaxies}
\label{sec:galprop}

The main data for the 9 probable satellite galaxies are presented in
Table 1, which is organised as follows.  Columns 1 and 2 give the UGC
number and classification of the central galaxy, with the latter being
taken from the NASA Extragalactic Database (NED henceforth). Column 3
gives the distance for the system, calculated from the central galaxy
recession velocity, using a Virgo-infall corrected model
\citep{paper1}.  Columns 4 and 5 give the \Ha -derived SF rate and
error for the central galaxy, and columns 6 and 7 the $R$-band total
magnitude and error.  The corresponding $R$-band luminosity in solar
units is given in column 8. Columns 9 and 10 give the SF rate and
error for the satellite galaxy, calculated as described in section 2,
and columns 11 and 12 the satellite galaxy $R$-band luminosity and
error.  The final column lists the projected separation in kpc between
the centres of the central and satellite galaxies.  This assumes that
central and satellite galaxies are at the same distance, as listed in
column 3.

Images of the 9 galaxy systems are shown in Appendix B. In each case,
the upper $R$-band image is a sufficently wide field to show both
central and satellite galaxies, and again lines have been added
  to indicate the location of the satellites.  The satellites are
shown in more detail in the lower images, with the $R$-band image on
the left and the continuum-subtracted \Ha\ image on the right.  In
every case, even the upper image shows only a subset of the area
searched for satellites.

Our method of identifying companions as likely satellites relies on
the detection of line emission in a narrow-band filter selected to
include \Ha\ from the central galaxy.  This method is not foolproof,
and it is possible that some of the putative satellites are background
galaxies with shorter-wavelength lines redshifted into the same
bandpass.  One check for this is to search for previously catalogued
recession velocities for the 9 satellites in the literature.  The
results of this search, using the NED ``Search for Objects Near Object Name''
facility, are given in Table 2.  All bar one had names or identifiers
in at least one catalogue or survey, listed in the second column of
table 2; the angular separation between central and satellite galaxies
is given in arcmin in column 3, and in kpc in column 4.  Only 3 satellites 
had separate recession velocities; these are listed in column 5, with the 
central galaxy recession
velocities being in the final column.  All 3 satellite velocities lie
within $\sim$100~km~s$^{-1}$ of that of the central galaxy, which gives 
some confidence in our methods.

Comments on individual galaxies:\\
{\bf UGC~2603} - the main galaxy is classified as an Im, so this can be 
considered a binary pair rather than a 
central plus satellite system.  The satellite galaxy has substantially 
higher surface brightness in $R$-band continuum than the \Ha GS galaxy, 
but fairly diffuse \Ha\ emission.\\
{\bf UGC~4273} - the companion is one of the most weakly detected in \Ha\,
but this has an independent recession velocity listed in NED, which is
very close to that of the central galaxy.\\
{\bf UGC~4362} - the companion is bright in the $R$-band image, and shows
strong SF with clumpy, resolved \HII\ regions, but has not been previously 
catalogued according to NED. The companion is on the extreme edge of the 
CCD frame so the measured SF rate and $R$-band luminosities in Table 1 
should be considered as lower limits, although the majority of the companion 
does appear to have been imaged, from inspection of sky survey 
images of this field.\\
{\bf UGC~4469} -  the companion is clearly a Magellanic irregular, with 
clumpy \Ha\ emission indicating several off-centre \HII\ regions. There
is an independent recession velocity for the companion in NED, again very
close to that of the the central galaxy.\\ 
{\bf UGC~4541} - this system appears to be undergoing significant
tidal disturbance; unsurprisingly, the catalogued recession velocity
of the companion is close to that of UGC~4541.  The \Ha\ morphology of
the companion reveals a strong nuclear starburst and/or AGN activity;
the strength of the \Ha\ emission results in the short SF timescale
found for this object in section 3.\\
{\bf UGC~4574} - the \Ha\ image of the companion shows two clumps of SF, 
located at either end of the moderately elongated $R$-band light 
distribution.\\
{\bf UGC~5688} - the central galaxy has low $R$-band luminosity and 
surface brightness.
The \Ha\ image shows clumps of emission from both ends of the elongated 
$R$-band light distribution, like the companion of UGC~4574, resembling 
the pattern of SF seen in many galaxy bars.\\
{\bf UGC~6506} - this is a similar case to UGC~2603 and UGC~5688, with a 
low luminosity central galaxy, and this should probably be considered a
binary system rather than as a central galaxy with a satellite.\\
{\bf UGC~12788} - the putative companion is projected on the outer
disk of the central galaxy, but does not look like an HII region or
part of a spiral arm.  This is a possible case of tidal interaction,
revealed in the distorted and strongly star-forming arm in UGC~12788,
close to the putative companion.  However, it should be noted that the
object considered here to be a companion galaxy is listed in NED as 
`West \HII\ region in UGC~12788'.\\

Two of the central/satellite systems studied by \citet{zari97} are in
the current sample.  The first is NGC~5921 and and its companion,
identified by \citet{zari97} as NGC~5921:[ZSF97]b. This companion is
clearly detected in our data, both in $R$-band and \Ha\ emission, and indeed was
independently selected as a putative satellite (it is the square at SF
rate $=$0.012 M$_{\odot}$ yr$^{-1}$, $R$-band luminosity 4 $\times
10^6$ L$_{\odot}$ in Fig. \ref{fig:dsfvdrt}).  However, this region,
along with one other similar region in NGC 5921 (not identified by
\citet{zari97}), was considered highly likely to be an outer
\HII\ region.  The second of the systems listed by \citet{zari97} is
NGC~5962 and the companion they identify as NGC~5962:[ZSF97]b. The
region containing the latter was searched in the present study, but
the companion is present only as a very faint $R$-band source, with no
detected \Ha\ emission.

The NED ``Search for Objects Near Object Name'' was also carried out for
the 23 outer disk objects.  None of these was found to a measured recession
velocity, and the only two to have individual NED entries are both classified 
as \HII\ regions, one in UGC~9935/NGC~5964 \citep{brad06}, and one in 
UGC~12343/NGC~7479 \citep{roza99}.

\begin{table*}
\begin{center} 
\begin{tabular}{rlcccc}
\hline
\hline
UGC$_{\rm c}$ &  Satellite name  &  Separation & Separation & 
V$_{\rm rec}$ sat  &  V$_{\rm rec}$ cent \cr
  &  & (arcmin) & (kpc) & (km~s$^{-1}$) & (km~s$^{-1}$) \cr
\hline
 2603 & 2MASXJ03192345+8116238 & 4.4 & 43.1 & -- & 2516 \\
 4273 & KUG0809+363 & 3.5 & 36.0 & 2483 & 2471 \\
 4362 &  -- & -- & -- & -- & 2344 \\
 4469 & NGC~2406B & 3.6 & 33.0 & 2104 & 2078 \\
 4541 & CGCG~060-036 & 2.4 & 21.9 & 2115 & 2060 \\
 4574 & 2MASXJ08482381+7402176 & 3.9 & 35.3 & -- & 2160 \\
 5688 & VV~294b & 1.9 & 16.1 & -- & 1920 \\
 6506 & MAPS-NGP O$\_$319$\_$1199567 & 3.6 & 30.5 & -- & 1580 \\
12788 & UM~007 NED01 & 1.0 & 9.5 & -- & 2956\\
\hline
\end{tabular}
\caption[]{Catalogue names and recession properties of previously-identified  
satellite galaxies}
\label{tbl:satname}
\end{center}
\end{table*}

\section{Statistics of star forming satellites}
\label{sec:galstats}

The next stage of the analysis is to look at the numbers of
star-forming satellites found, to put constraints on the overall
abundance of such systems around field galaxies.
Given that the regions around
119 \Ha GS galaxies were searched, 9 satellites found seems 
a small number, given that we have two such systems around the 
Milky Way, but in order to make this comparison quantitative 
we need to correct for sources of incompleteness in our search method.

\begin{figure}
\centering
\rotatebox{-90}{
\includegraphics[height=8.5cm]{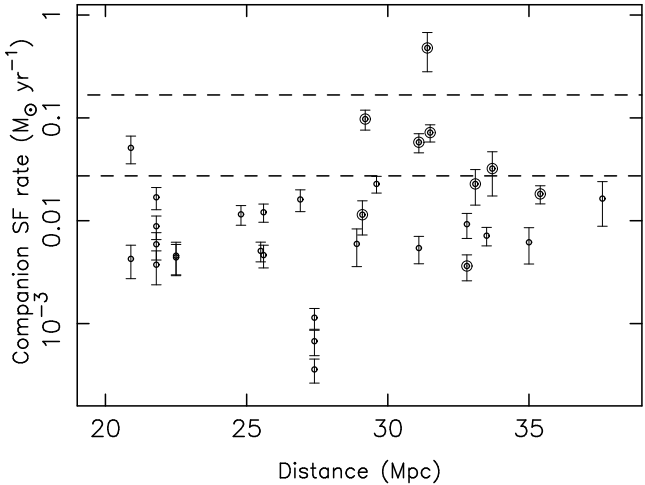}
}
\caption{
Companion object SF rate vs the distance of the central
object, in Mpc.  The dashed lines show the SF rates corresponding to the
LMC (upper line) and SMC (lower line), demonstrating that we can detect
companions with SF rates much lower than those of the Magellanic Clouds
to the limits of our survey.  
The ringed points indicate the 9 probable
satellite galaxies, as explained in section 3.
}
\label{fig:dsfvdmpc}
\end{figure}

As a starting point for this analysis, we need to confirm that our
data and methods are sufficiently sensitive to detect `Magellanic
Cloud like' satellites.  Figure \ref{fig:dsfvdmpc} shows the \Ha
-derived SF rates for 32 regions identified in the current paper as
possible satellites, with the ringed points identifying the 9 likely
satellites.  These rates are plotted against the distance in Mpc of
the central galaxy in the system, and the horizontal lines show the SF
rates of the Magellanic Clouds.  This confirms that our H$\alpha$
technique is easily sensitive enough to detect star forming galaxies
fainter than the Magellanic Clouds to the spatial limits of our
survey, as objects with \Ha\ luminosities significantly lower than
that of the SMC are seen over the full range of distances studied.

\begin{figure}
\centering
\rotatebox{-90}{
\includegraphics[height=8.5cm]{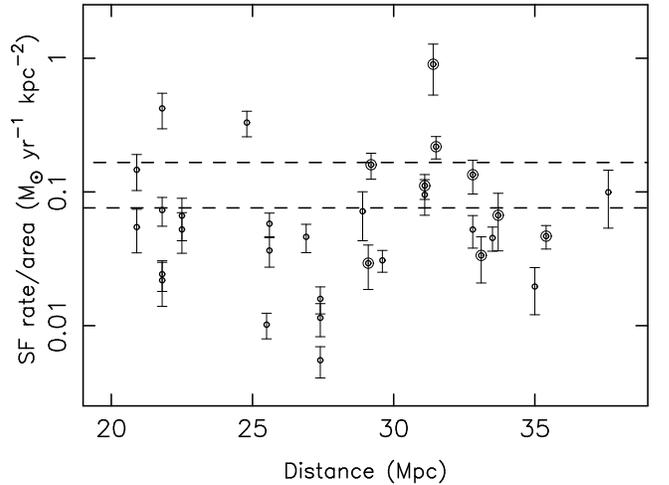}
}
\caption{
Companion object SF rate per unit area vs the distance of the central
object, in Mpc.  The dashed lines show the SF rates per unit area 
corresponding to the
LMC (upper line) and SMC (lower line), demonstrating that we can detect
companions with SF surface densities much lower than those of the 
Magellanic Clouds
to the limits of our survey.  
The ringed points indicate the 9 probable
satellite galaxies, as explained in section 3.
}
\label{fig:dsbvdmpc}
\end{figure}

Surface brightness is also important in determining the detectability
of galaxies; it is possible that Magellanic-type companions with quite
high total SF rates could be missed if the line emission were very
extended and hence of low surface brightness.  This possibility is
investigated in Fig. \ref{fig:dsbvdmpc}, which is similar to
Fig. \ref{fig:dsfvdmpc}, but with the SF rate of each region divided
by its area in kpc$^2$, measured from our \Ha\ images.  The same
quantity for the Magellanic Clouds is again shown by the dashed lines.
The latter values were derived from \citet{kenn86}, who give \Ha\ flux
measurements measured in large apertures comparable in kpc sizes to
the apertures used to measure the \Ha\ fluxes in our companion
objects.  Specifically, \citet{kenn86} find 25\% of the \Ha\ flux from
the LMC to come from a 30~arcmin aperture centred on 30 Doradus; this
would be both spatially resolved at the distance of our satellite
galaxies, and easily detectable in surface brightness as shown from
the position of the upper dashed line in Fig. \ref{fig:dsbvdmpc}.
Similarly, the lower dashed line comes from a 16~arcmin aperture flux
measurement on the SMC, which contains 17\% of the total \Ha\ emission
and hence of the inferred SF activity in that galaxy.  Thus we can
again conclude that Magellanic-type galaxies should be easily
detectable with our data and techniques.

In order to calculate incompleteness corrections, it is necessary to
decide on a definition of a close companion galaxy.  For the present
work, a `Magellanic-type' companion is defined to be one that lies
within a spherical region, centred on the UGC galaxy, with a volume
twice as large as the volume that just contains both the Magellanic
Clouds. This corresponds to a radial separation of less than $60
\times \sqrt[3] 2 = 75.2$kpc, where 60~kpc is our adopted distance to
the SMC.  However, for any given galaxy, it is not possible to see the
full volume potentially occupied by satellites.  For closer galaxies,
the outlying parts of the volume are missed as they lie outside the
CCD field of view. In addition, there is a `blank spot' along the
line-of-sight of the main galaxy, in all cases, as the \Ha\ emission
from any companion would be confused with that of the central galaxy,
in our narrow-band images.  Knowing the distance of each of the
central galaxies, their optical major and minor axes, and the size of
the imaged field, it is then possible to calculate the fraction of the
spherical halo volume that is missed due to these two causes. This fraction
was then assumed to be equal to the fraction of satellites that were
missed, i.e. implicitly assuming that satellites were equally likely
to be found anywhere within the spherical volume.  In support of this
assumption, several studies have found satellite distributions to be
isotropic, for blue central galaxies such as those studied here
\citep{yang06,agus07,azza07}.

A programme was written to calculate the completeness fraction for all
119 of the \Ha GS galaxies, and the results are summarised in Table 
\ref{tbl:satstats}.  This lists, as a function of Hubble $T$-type, the 
total numbers of galaxies observed (column 2); the sums of the observable 
fractions of the satellite volumes (column 3); the number of star-forming
satellites found (column 4); and ratio of satellites found to observable 
volumes, i.e. column 4 divided by column 3 (listed in the final column).
To summarise these results, the 119 potentially satellite-containing volumes 
were covered with a mean efficiency of just below 50\%, resulting in an 
effective search of 53 volumes.  This search yielded 9 star-forming 
satellites, or 0.17$^{+0.08}_{-0.06}$ satellite galaxies per full volume searched.
This fraction applies whether all T-types included, or just the more
luminous galaxies with $T$ types in the range 0 to 5 (bottom and penultimate 
lines of Table \ref{tbl:satstats}).

The errors quoted here and in the final column of Table \ref{tbl:satstats}
are 1-$\sigma$ limits derived from the tables
of \citet{gehr86}.

It could be argued that the companion to UGC~12788, which lies inside the 
projected disk of UGC~12788 but is detected due to its distinctive 
morphology, should be excluded from these statistics.  It should also be 
noted that some of the detected satellites may actually lie outside the
75.2~kpc radius circle due to projection uncertainties.  Thus the fractions
listed in the final column of Table \ref{tbl:satstats} could be 
considered as upper limits for the numbers of `Magellanic Cloud type' 
companions as we have defined them here. However, we should also take
into account to possibility that some of the objects classified as outlying
\HII\ regions may in fact be satellite galaxies.  If the two `possible' objects
identified from their $R$-band sizes in Fig. \ref{fig:Rsatsize} are included,
the overall fraction of satellites per central galaxy search increases from 
0.17 to 0.21$^{+0.08}_{-0.06}$.  If we very conservatively include all the 
outlying objects not positively identified elsewhere as being \HII\ regions, 
the total number increases to 30, for an overall fraction of satellites per
galaxy searched of 0.57$^{+0.12}_{-0.10}$.

\begin{table*}
\begin{center} 
\begin{tabular}{crrcc}
\hline
\hline
$T$-type &  N$_{\rm Gal}$  &  N$_{\rm Corr}$ &  N$_{\rm Sat}$  &  
N$_{\rm Sat}$/N$_{\rm Corr}$  \cr
\hline
  0  &  6  &  2.4  &  1  &  0.42$^{+0.97}_{-0.35}$ \\
  1  &  7  &  3.5  &  1  &  0.29$^{+0.67}_{-0.24}$ \\
  2  &  4  &  2.1  &  0  &  0.00$^{+0.88}_{-0.00}$ \\
  3  & 15  &  7.2  &  2  &  0.28$^{+0.37}_{-0.18}$ \\
  4  & 15  &  6.8  &  0  &  0.00$^{+0.27}_{-0.00}$ \\
  5  & 14  &  5.9  &  1  &  0.17$^{+0.39}_{-0.14}$ \\
  6  & 15  &  6.5  &  1  &  0.15$^{+0.35}_{-0.12}$ \\
  7  & 14  &  5.7  &  1  &  0.18$^{+0.41}_{-0.15}$ \\
  8  & 13  &  5.5  &  0  &  0.00$^{+0.33}_{-0.00}$ \\
  9  &  6  &  2.4  &  1  &  0.42$^{+0.97}_{-0.35}$ \\
 10  & 10  &  4.6  &  1  &  0.22$^{+0.51}_{-0.18}$ \\
\hline
0-5  & 61  & 28.0  &  5  &  0.18$^{+0.12}_{-0.08}$ \\
0-10 &119  & 52.7  &  9  &  0.17$^{+0.08}_{-0.06}$ \\
\hline
\end{tabular}
\caption[]{The number of central galaxies studied, broken down by Hubble type,
and the frequency with which they host star-forming satellites.}
\label{tbl:satstats}
\end{center}
\end{table*}

\section{Correlation of satellite and central galaxy properties}
\label{sec:galcorr}

The search carried out in the present study has identified 9 probable
satellite galaxies, with significant stellar masses, in close
proximity (in projection) to their central galaxies. Given these properties,
it is interesting to check whether the SF activity, either of the central 
galaxies or of the probable companions, is affected by the tidal forces 
between central and satellite galaxies.  The overall question of the effect
of environment on SF rates for the entire \Ha GS sample is the subject of a
future paper, so this analysis concerns only the subsample of galaxies
with identified companions.

\begin{figure}
\centering
\rotatebox{-90}{
\includegraphics[height=8.5cm]{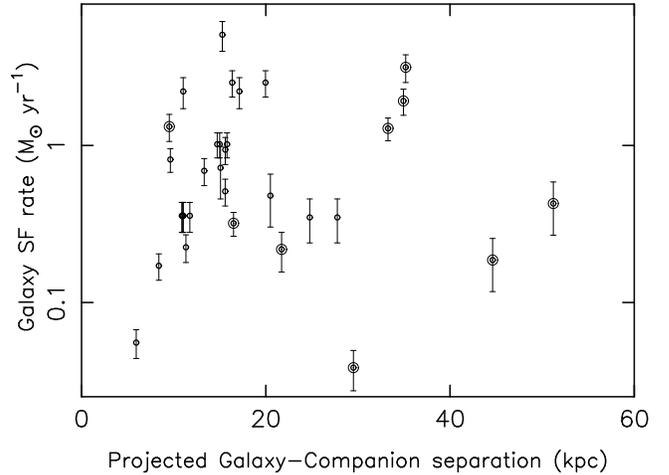}
}
\caption{
The central galaxy SF rate plotted against the projected separation 
of the companion object and the central galaxy.
The ringed points indicate the 9 probable
satellite galaxies, as explained in section 3.
}
\label{fig:gsfvdgd}
\end{figure}

\begin{figure}
\centering
\rotatebox{-90}{
\includegraphics[height=8.5cm]{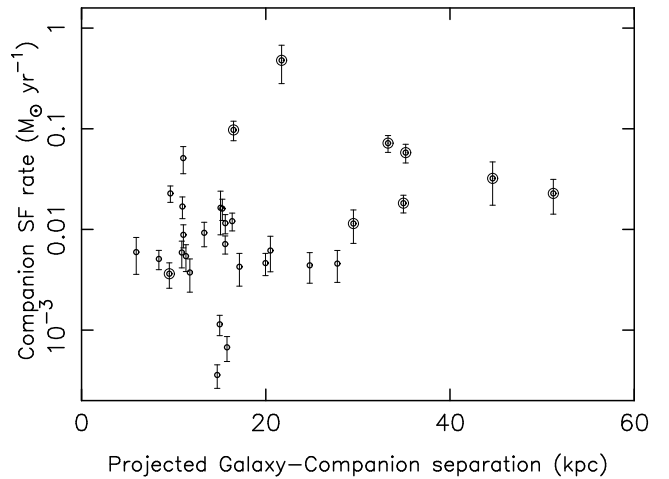}
}
\caption{
The SF rate for each companion object plotted against the projected 
separation of the companion
and the central galaxy.
The ringed points indicate the 9 probable
satellite galaxies, as explained in section 3.
}
\label{fig:dsfvdgd}
\end{figure}

Figure \ref{fig:gsfvdgd} shows the \Ha -derived SF rates of the central 
galaxies plotted against projected galaxy-companion separation in kpc.
The ringed points indicate the 9 probable satellite systems as identified
in section 3.  This figure shows no correlation between central galaxy SF
rate and projected separation, so there is no obvious effect of the 
presence of the satellites on the their central galaxies. Similarly,
Fig. \ref{fig:dsfvdgd} shows no correlation between satellite galaxy SF rate 
and projected distance from the central galaxy, at least for the 9 probable
satellites.  There is a significant difference between these probable satellites
and the objects identified in section 3 as outlying \HII\ regions, but this is
as expected under our preferred interpretation of these objects.

\begin{figure}
\centering
\rotatebox{-90}{
\includegraphics[height=8.5cm]{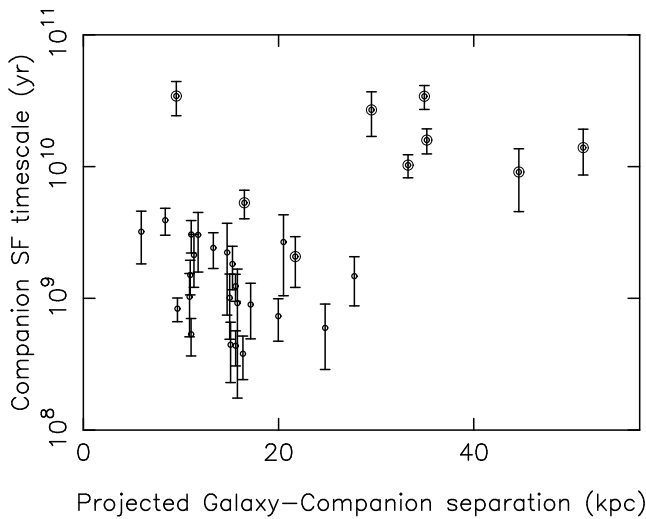}
}
\caption{
The SF timescale of each companion object plotted against the projected 
separation of the companion and the central galaxy.
The ringed points indicate the 9 probable
satellite galaxies, as explained in section 3.
}
\label{fig:dsftvdgd}
\end{figure}

Finally, Fig. \ref{fig:dsftvdgd} shows companion object SF timescales,
as defined in section 3, against projected separation between central
and satellite galaxies.  Again, the only clear result from this plot
is the short SF timescales of the outlying \HII\ regions, already
identified in Fig. \ref{fig:sftsat}. For the 9 probable satellites, no
correlation is found between SF timescale and projected separation
from their central galaxies.  The satellite galaxy with the shortest
SF timescale is the companion to UGC~4541; this not the closest
galaxy-satellite pair investigated here, but there does appear to be
tidal distortion associated with this interaction.  The closest
pairing in projected separation is UGC~12788 and its companion; here
the companion appears completely unaffected by tidal effects, in terms
of both optical morphology and SF properties, so it is possible that
there is a significant line-of-sight separation in this case.  The
second closest in apparent separation are UGC~5688 and companion; the
latter has a SF timescale below 10~Gyr, which may show a modest
enhancement in SF rate as a the result of tidal effects.  However,
UGC~5688 is a low luminosity galaxy of type SBm, and so tidal effects
are likely to be minimal due to the low mass of the primary galaxy.

\section{Discussion and future work}
\label{sec:disc}

The major result of this paper is that actively star-forming
  satellite galaxies, with luminosities and SF rates within
  approximately an order of magnitude of those of the Magellanic
  Clouds, are fairly rare.  Most galaxies in our survey of field
  spirals have no such satellites.  Previous searches for satellites
  of all types \citep{zari97, mado04} have found significantly more
  than this, typically one substantial satellite per central galaxy
  surveyed, although it is very difficult to make quantitative
  comparisons because of the varying depths and methods of the
  different surveys.  However, it does appear likely that most
  satellites, even of isolated field galaxies, are of non-star forming
  types.  Examples close to home are M~32 and NGC~205, early type
  companions of the Andromeda spiral M~31.  Given that most isolated
  galaxies of Magellanic-Cloud like luminosities are actively forming
  stars (paper V), this would seem to imply the efficient
  truncation of star formation in satellites,
  with a timescale short compared to a Hubble time.  This then raises
  interesting questions about the fate of the gas originally in the
  satellite dwarfs; is it consumed in a burst of star formation
  triggered by interaction with the central galaxy, or is a
  significant fraction of it expelled from the satellite, enabling it
  to be accreted onto the central galaxy? Evidence of this process in
  the Milky Way system may be provided by the Magellanic Stream.  
Given the importance of gas supply in disk galaxy evolution, it would
clearly be useful to know the frequency of such gas-yielding interactions,
and the timescales and gas masses involved.

Observationally, there are 3 main requirements: to extend the sample size of 
star-forming satellite candidates; to discriminate definitively between 
true satellites and outlying parts of disks; and to determine the red-to-blue
satellite fraction.  The first can simply be done by further \Ha\ imaging, 
preferably going rather deeper and over wider fields than the present data set.
The second requires spectroscopy to determine velocity differences 
between central and putative satellite galaxies, and preferably velocity
cubes (e.g. from \Ha\ Fabry-Perot instruments or \HI\ maps) to look for
signs of interaction.  Data enabling the identification of quiescent 
red-sequence satellites are problematic given the greater difficulty of 
measuring recession velocities for faint and possibly low-surface-brightness
absorption line sources, but the situation is being improved with surveys
using sensitive
multi-fibre instruments.  We are actively pursuing all of these approaches.

One caveat on the current results concerns the conclusion that the 9
satellites are fairly similar to the Magellanic Clouds.  This can be seen
as unsurprising given that we used the similarity to the Magellanic Clouds,
particularly in Figs. \ref{fig:dsfvdrt} and \ref{fig:sftsat}, to argue
for the 9 objects being true satellites.  There is indeed circularity in this
argument, and it is clearly important to include the provisionally rejected
objects in follow-up spectroscopy.
Given that the latter tend to be found at small projected separations
from the central galaxies and have similar properties to \HII\ regions in 
Fig. \ref{fig:sftsat}, it is likely that this interpretation will be confirmed 
for most or all of them, but there may be some important objects lurking in 
this category.

\section{Conclusions}
\label{sec:conc}

This study has identified 9 probable star-forming satellite galaxies
with projected separations consistent with their being as close to
their central galaxies as the Magellanic Clouds are to the Milky Way.
Figure \ref{fig:dsfvdrt} illustrates that the stellar luminosities and
SF rates of the Magellanic Clouds are comparable to those of the 9
probable satellites found here.  
Overall, the satellite galaxies (including the Magellanic Clouds) are
currently forming stars at a rate comparable to field Sm and Im
galaxies in the \Ha GS sample. The only evidence of a strong
starburst is in the tidally-disturbed companion to UGC~4541.
Considering the 9 probable satellites and the Magellanic Clouds
together, the LMC and SMC are the brightest and 7th brightest in
$R$-band luminosity, and the 2nd and 9th most rapidly star-forming.
Thus the LMC is clearly a large satellite, whereas the SMC is close to
or just below average amongst those found here.  We find no cases of 2
satellites around any of the 119 central galaxies studied, so the
Milky Way appears well-favoured in the number of large star-forming
companions in its immediate neighbourhood.
In this context, it is interesting to note that \citet{puec07} have 
recently concluded that the MW is distinctive for its 
{\em lack} of merging activity over its history; the results found here
indicate that any such deficiency is likely to be rectified in the future.

\begin{acknowledgements}
The Jacobus Kapteyn Telescope was operated on the island of La Palma
by the Isaac Newton Group in the Spanish Observatorio del Roque de los
Muchachos of the Instituto de Astrofisica de Canarias. This research
has made extensive use of the NASA/IPAC Extragalactic Database (NED)
which is operated by the Jet Propulsion Laboratory, California
Institute of Technology, under contract with the National Aeronautics
and Space Administration.  Jane O'Neill contributed to this work while 
on a summer placement supported by the Nuffield Science Bursary Scheme.
PAJ thanks Chris Moss and Sue Percival for useful comments, and the 
referee is also thanked for many constructive suggestions.
\end{acknowledgements}
\bibliographystyle{bibtex/aa}
\bibliography{refs}
      

\newpage
\begin{appendix}
\section{Images of the 2 possible satellites and their central galaxies}

\begin{figure}
\begin{center}
\includegraphics[height=4.0cm, width=4.0cm]{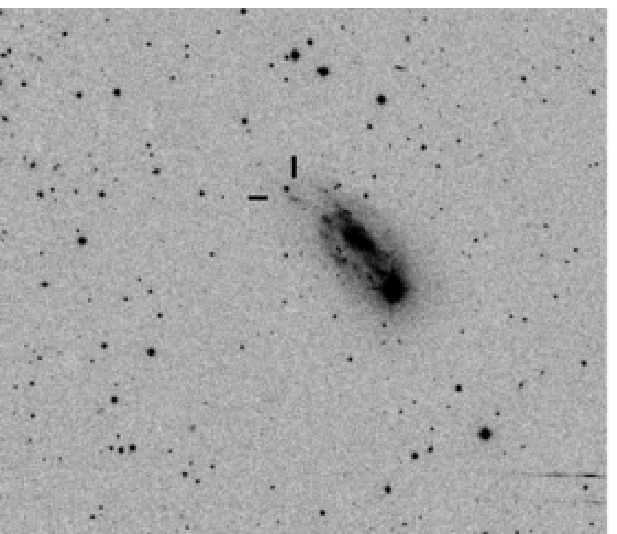}
\includegraphics[height=4.0cm, width=4.0cm]{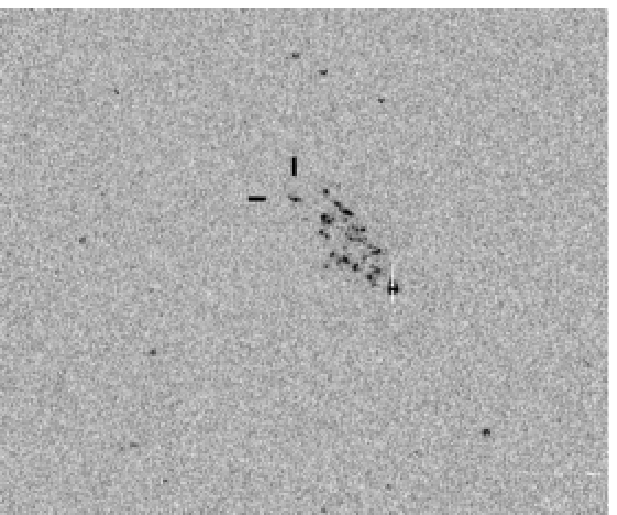}

\vspace*{0.4cm}
\includegraphics[height=4.0cm, width=4.0cm]{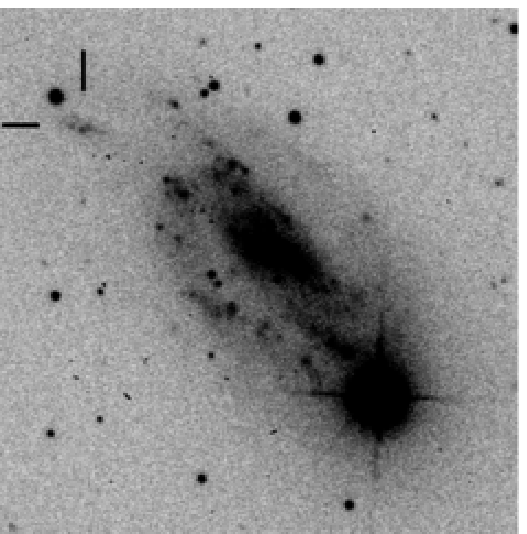}
\includegraphics[height=4.0cm, width=4.0cm]{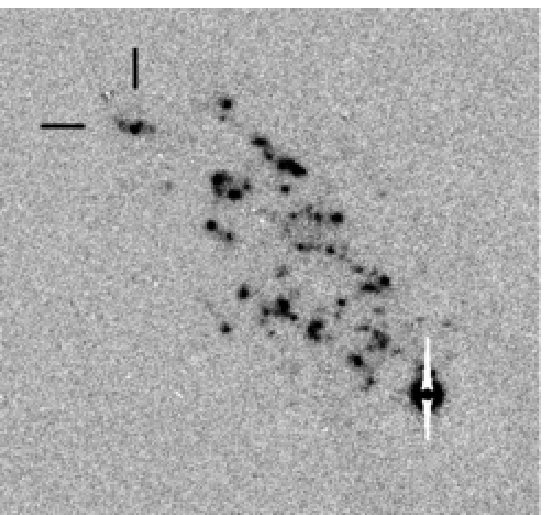}
\end{center}
\caption{
Upper images: UGC~3530 and satellite galaxy, showing the full field of view
(image size 160$^{\prime \prime}$ by 160$^{\prime \prime}$);
$R$-band (left) and \Ha\ (right).
Lower image: UGC~3530 and satellite galaxy, close-up view
(image size 160$^{\prime \prime}$ by 160$^{\prime \prime}$);
$R$-band (left) and \Ha\ (right).
}
\label{fig:u3530}
\end{figure}

\vspace*{1.0 cm}

\begin{figure}
\begin{center}
\includegraphics[height=4.0cm, width=4.0cm]{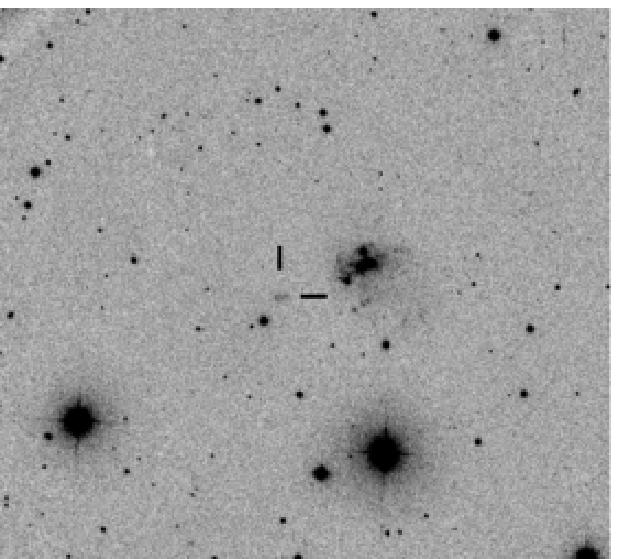}
\includegraphics[height=4.0cm, width=4.0cm]{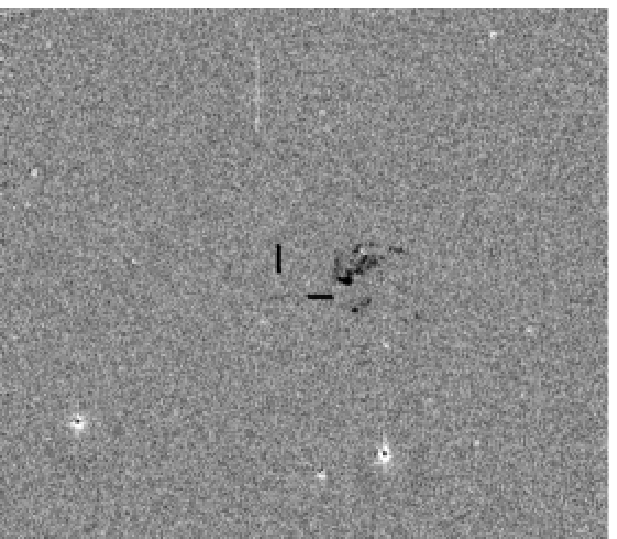}

\vspace*{0.4cm}
\includegraphics[height=4.0cm, width=4.0cm]{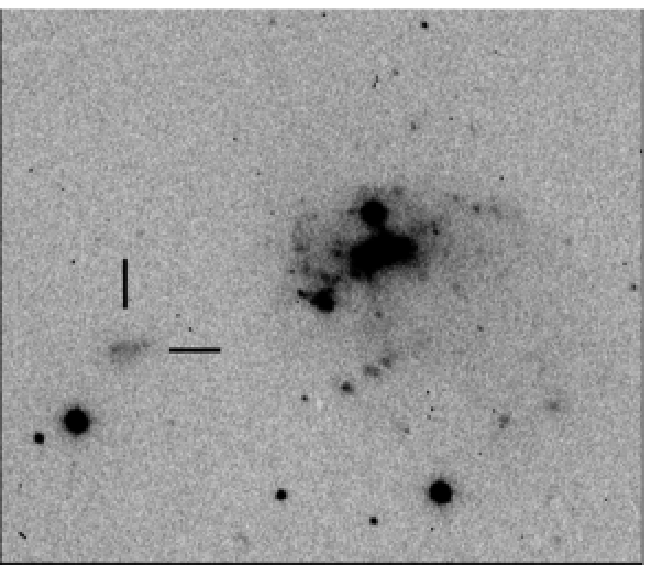}
\includegraphics[height=4.0cm, width=4.0cm]{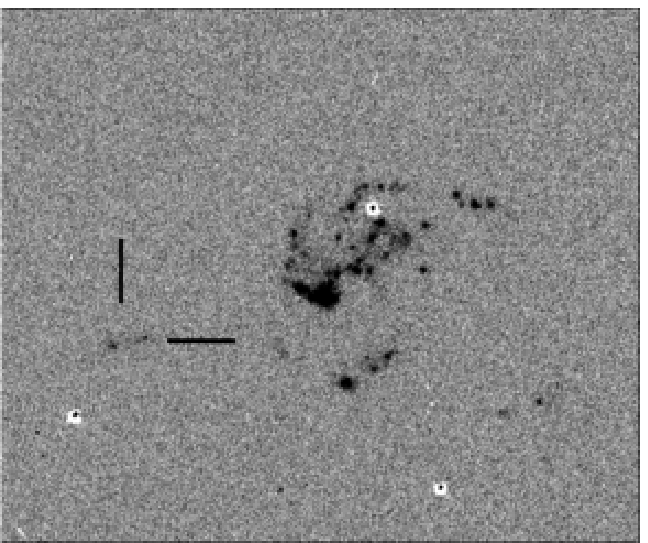}
\end{center}
\caption{
Upper images: UGC~4260 and satellite galaxy, showing the full field of view
(image size 560$^{\prime \prime}$ by 560$^{\prime \prime}$);
$R$-band (left) and \Ha\ (right).
Lower images: UGC~4260 and satellite galaxy, close-up view
(image size 154$^{\prime \prime}$ by 154$^{\prime \prime}$);
$R$-band (left) and \Ha\ (right).
}
\label{fig:u4260}
\end{figure}

\end{appendix}

\newpage
\begin{appendix}
\section{Images of the 9 probable satellites and their central galaxies}

\begin{figure}
\begin{center}
\includegraphics[height=4.0cm]{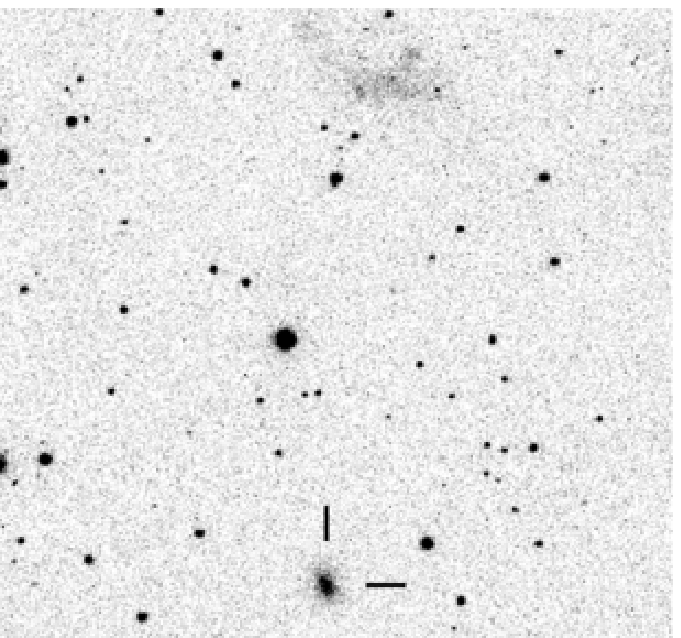}\\

\vspace*{0.4cm}
\includegraphics[height=4.0cm]{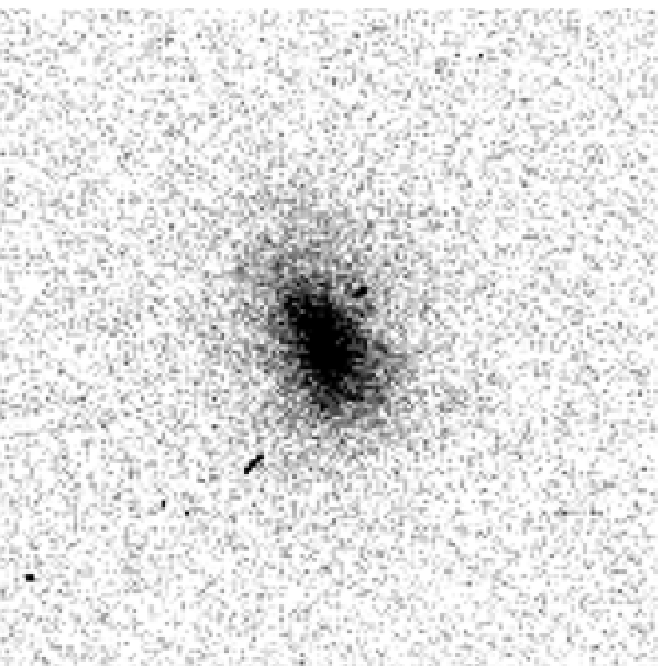}
\includegraphics[height=4.0cm]{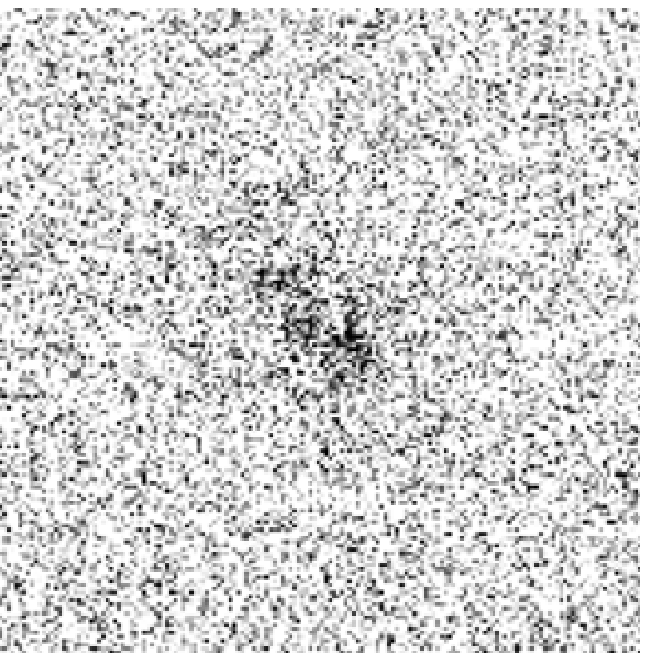}
\end{center}
\caption{
Upper image: UGC~2603 and satellite galaxy 
(image size 343$^{\prime \prime}$ by 325$^{\prime \prime}$).
Lower images: Satellite galaxy in $R$-band (left) and continuum-subtracted 
\Ha\ (right: image sizes 47$^{\prime \prime}$ by 47$^{\prime \prime}$).
}
\label{fig:u2603}
\end{figure}

\vspace*{1.0 cm}

\begin{figure}
\begin{center}
\includegraphics[height=4.0cm]{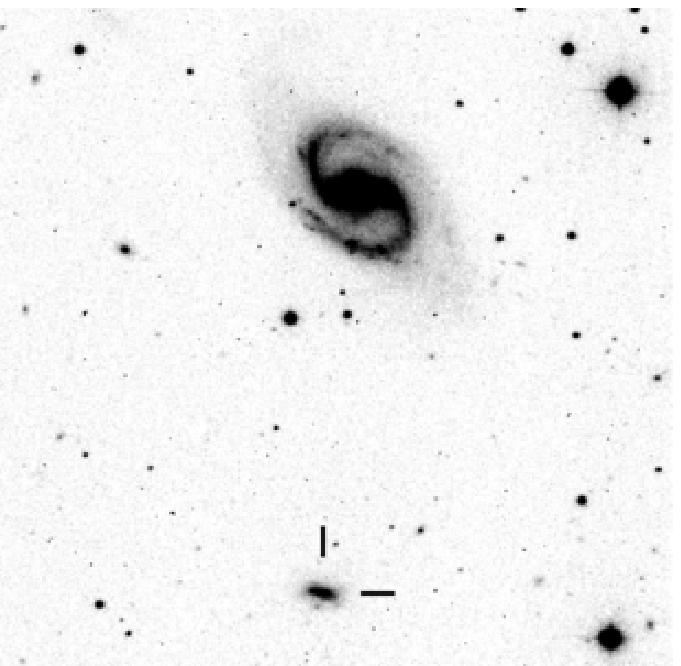}\\

\vspace{0.4cm}
\includegraphics[height=4.0cm]{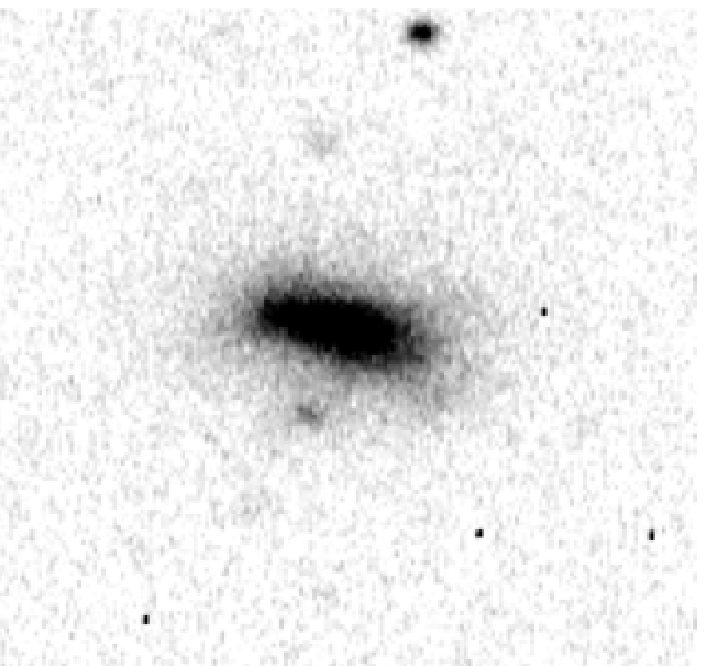}
\includegraphics[height=4.0cm]{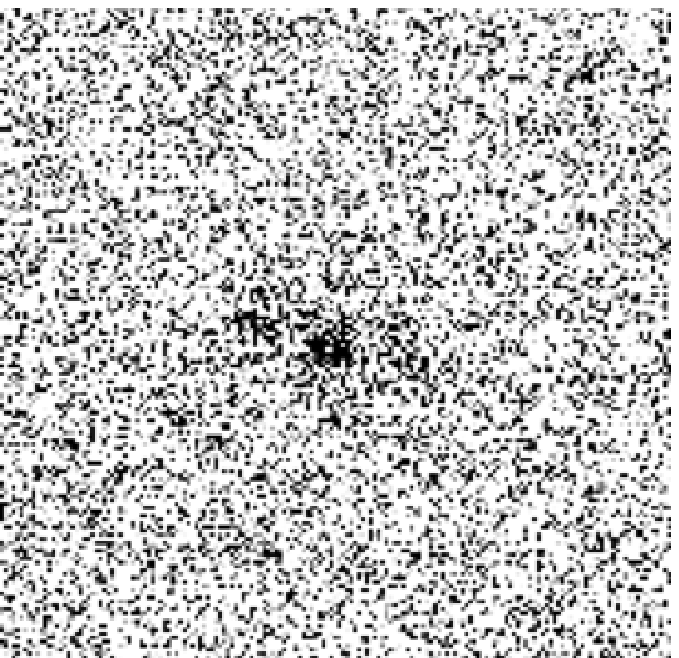}
\end{center}
\caption{
Upper image: UGC~4273 and satellite galaxy (image size 338$^{\prime \prime}$ by
335$^{\prime \prime}$).
Lower images: Satellite galaxy in $R$-band (left) and continuum-subtracted 
\Ha\ (right: image sizes 56$^{\prime \prime}$ by 56$^{\prime \prime}$).
}
\label{fig:u4273}
\end{figure}

\vspace*{1.0 cm}

\begin{figure}
\begin{center}
\includegraphics[height=4.0cm]{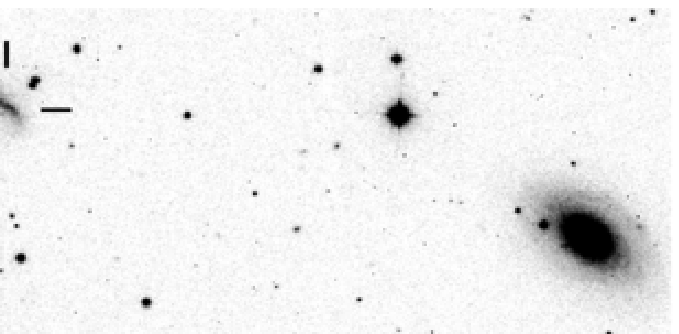}

\vspace*{0.4cm}
\includegraphics[height=4.0cm]{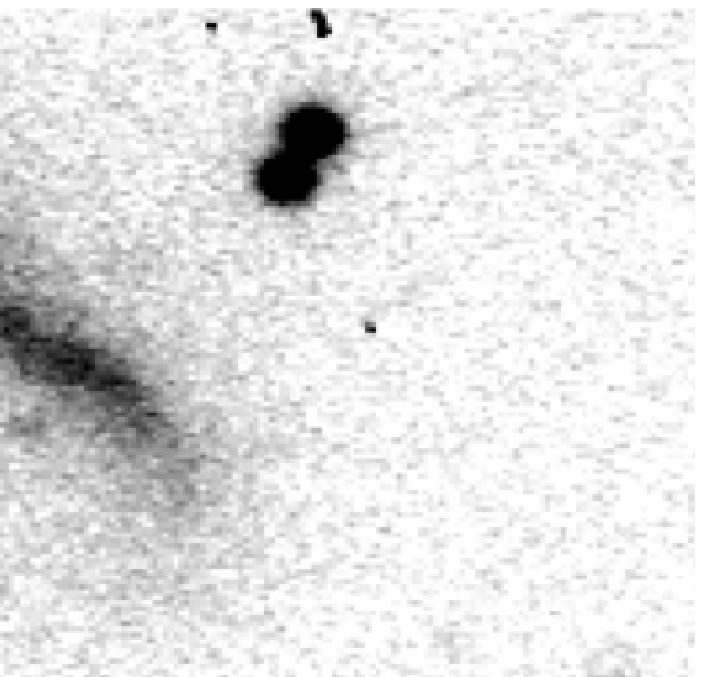}
\includegraphics[height=4.0cm]{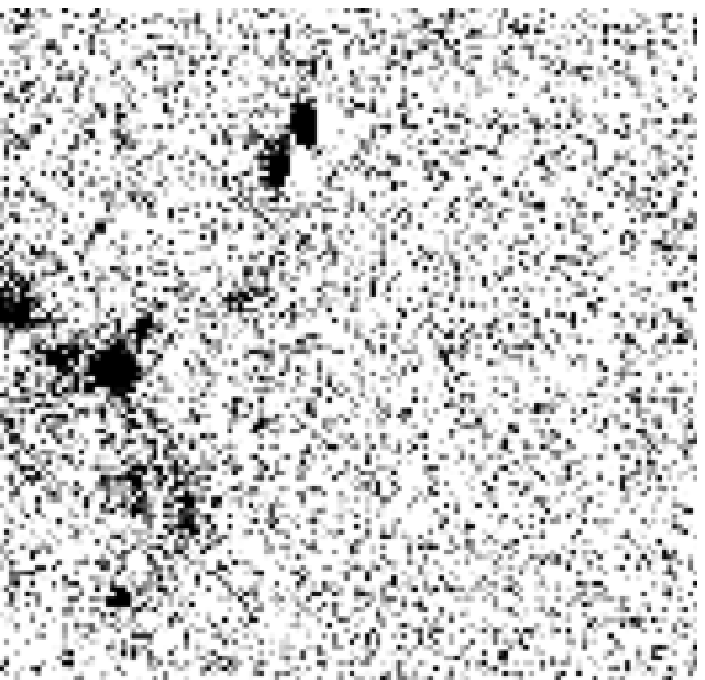}
\end{center}
\caption{
Upper image: UGC~4362 and satellite galaxy (image size 359$^{\prime \prime}$ by
177$^{\prime \prime}$).
Lower images: Satellite galaxy in $R$-band (left) and continuum-subtracted 
\Ha\ (right: image sizes 44$^{\prime \prime}$ by 44$^{\prime \prime}$).
}
\label{fig:u4362}
\end{figure}

\vspace*{1.0 cm}

\begin{figure}
\begin{center}
\includegraphics[height=4.0cm]{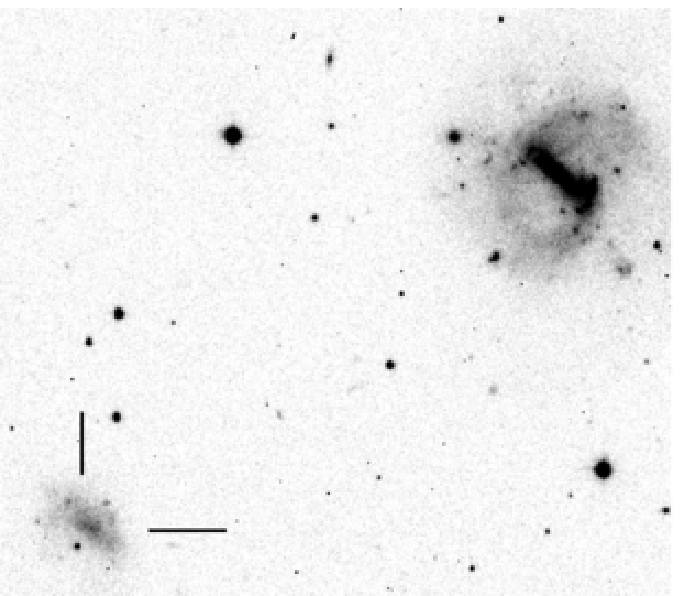}

\vspace*{0.4cm}
\includegraphics[height=4.0cm]{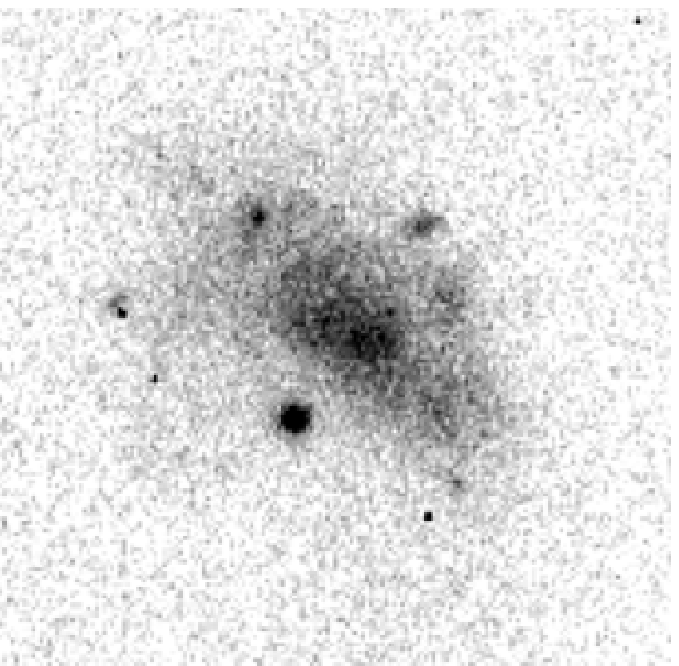}
\includegraphics[height=4.0cm]{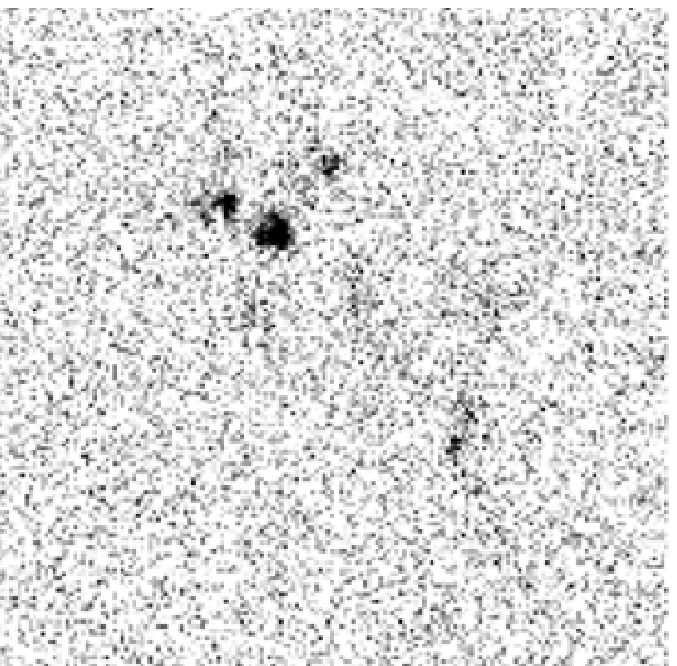}
\end{center}
\caption{
Upper image: UGC~4469 and satellite galaxy (image size 248$^{\prime \prime}$ by
220$^{\prime \prime}$).
Lower images: Satellite galaxy in $R$-band (left) and continuum-subtracted 
\Ha\ (right: image sizes 51$^{\prime \prime}$ by 51$^{\prime \prime}$).
}
\label{fig:u4469}
\end{figure}

\vspace*{1.0 cm}

\begin{figure}
\begin{center}
\includegraphics[height=4.0cm]{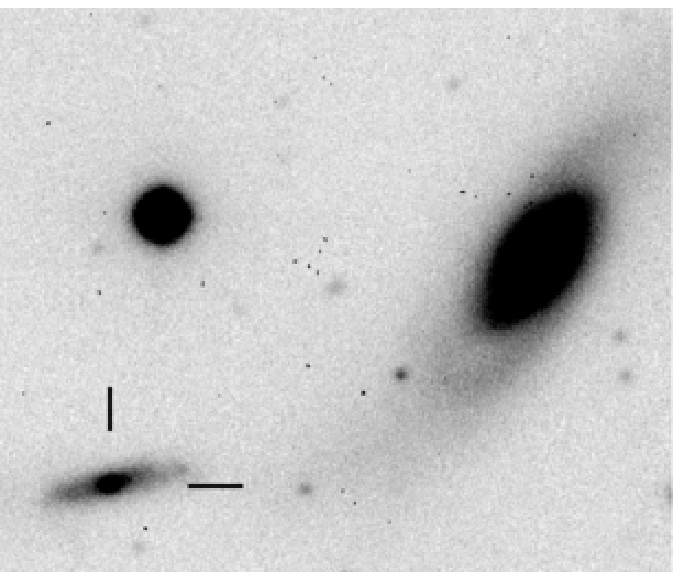}

\vspace*{0.4cm}
\includegraphics[height=4.0cm]{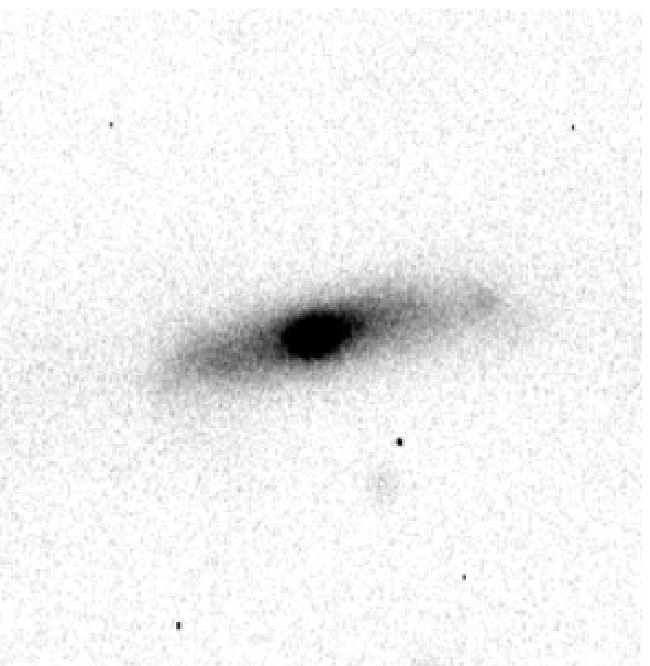}
\includegraphics[height=4.0cm]{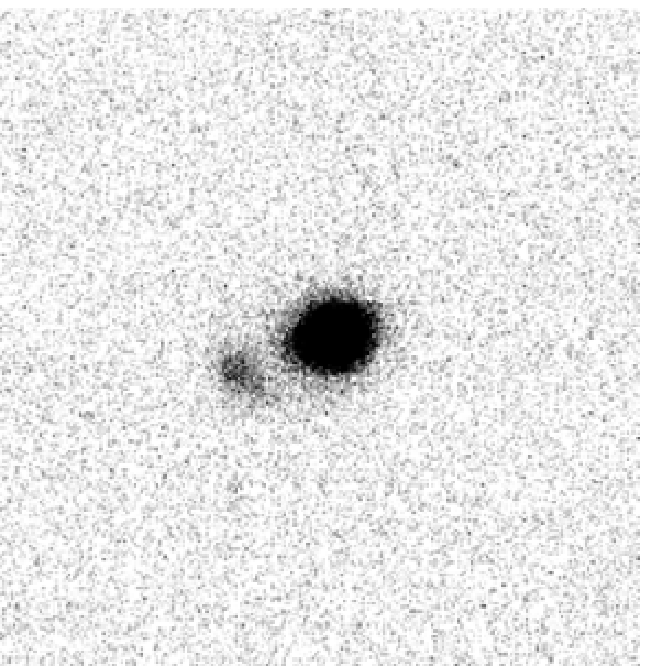}
\end{center}
\caption{
Upper image: UGC~4541 and satellite galaxy (image size 196$^{\prime \prime}$ by
161$^{\prime \prime}$).
Lower images: Satellite galaxy in $R$-band (left) and continuum-subtracted 
\Ha\ (right: image sizes 78$^{\prime \prime}$ by 78$^{\prime \prime}$).
}
\label{fig:u4541}
\end{figure}


\begin{figure}
\begin{center}
\includegraphics[height=4.0cm]{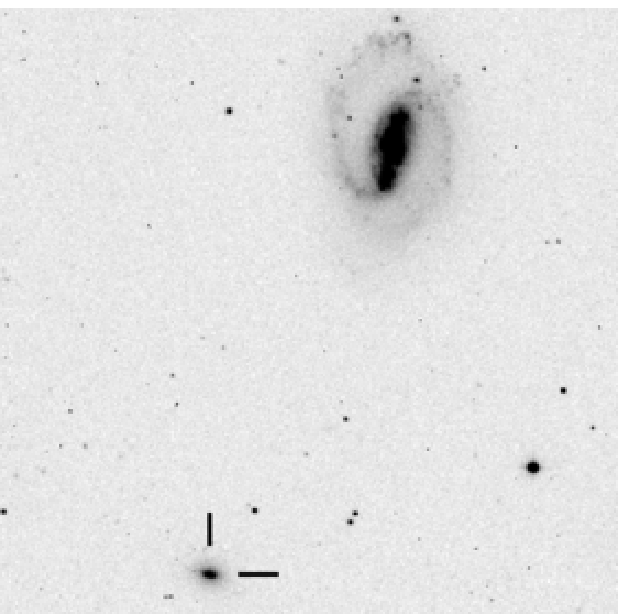}

\vspace*{0.4cm}
\includegraphics[height=4.0cm]{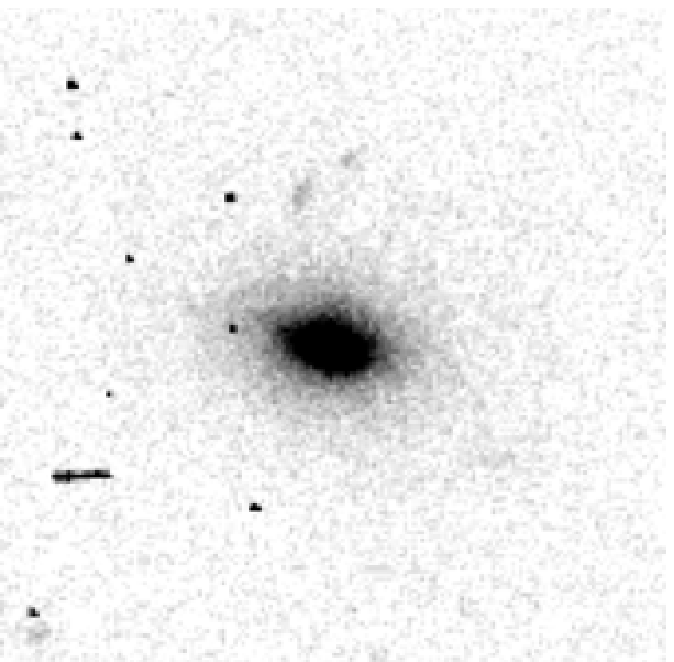}
\includegraphics[height=4.0cm]{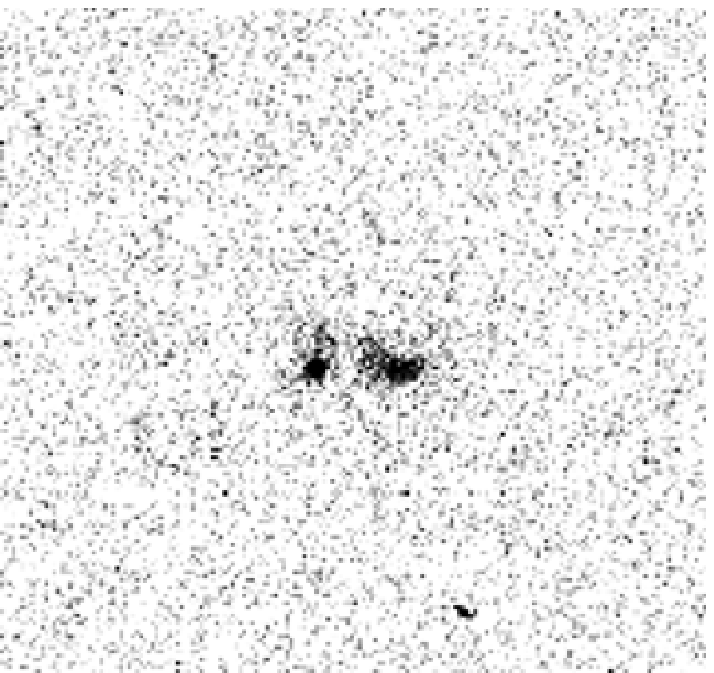}
\end{center}
\caption{
Upper image: UGC~4574 and satellite galaxy (image size 318$^{\prime \prime}$ by
309$^{\prime \prime}$).
Lower images: Satellite galaxy in $R$-band (left) and continuum-subtracted 
\Ha\ (right: image sizes 53$^{\prime \prime}$ by 53$^{\prime \prime}$).
}
\label{fig:u4574}
\end{figure}

\vspace*{1.0 cm}

\begin{figure}
\begin{center}
\includegraphics[height=4.0cm]{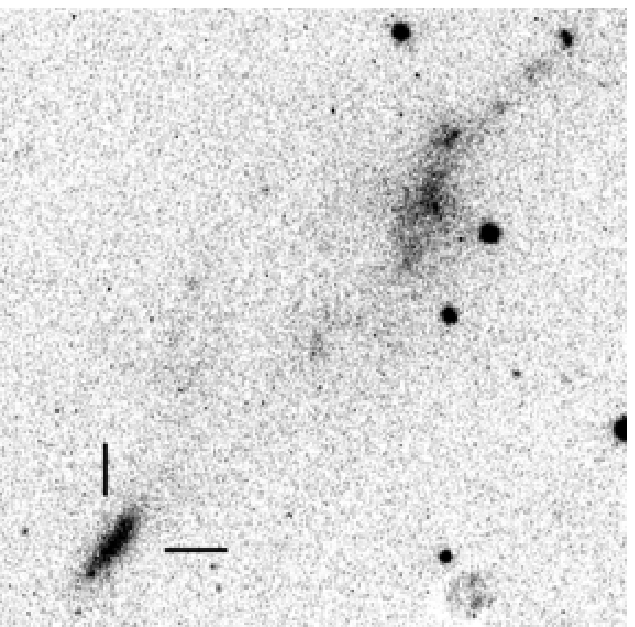}

\vspace*{0.4cm}
\includegraphics[height=4.0cm]{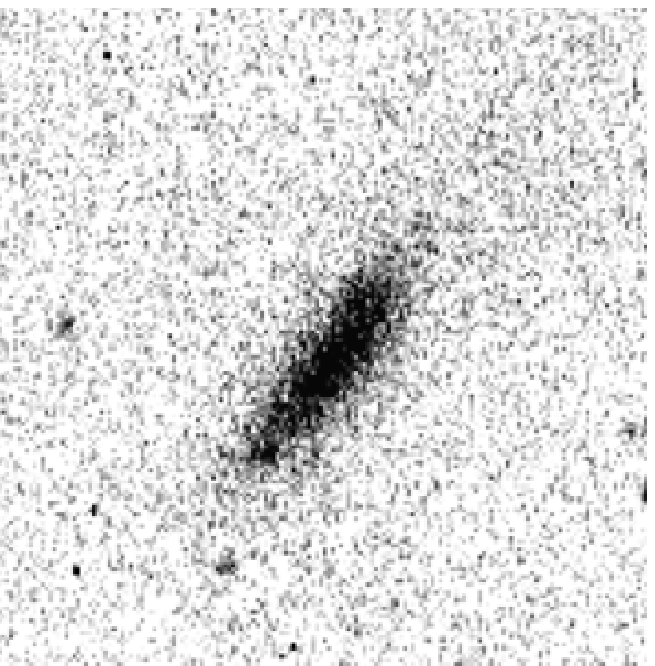}
\includegraphics[height=4.0cm]{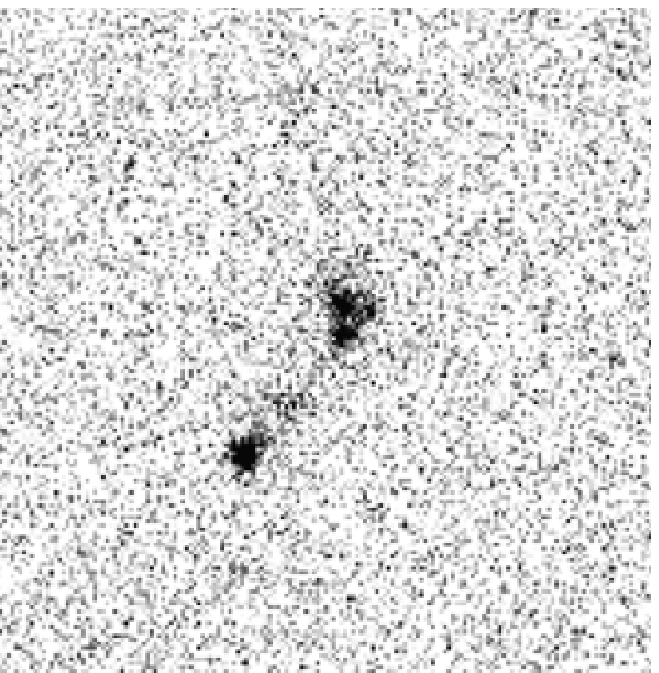}
\end{center}
\caption{
Upper image: UGC~5688 and satellite galaxy (image size 160$^{\prime \prime}$ by
154$^{\prime \prime}$).
Lower images: Satellite galaxy in $R$-band (left) and continuum-subtracted 
\Ha\ (right: image sizes 51$^{\prime \prime}$ by 51$^{\prime \prime}$).
}
\label{fig:u5688}
\end{figure}

\vspace*{1.0 cm}

\begin{figure}
\begin{center}
\includegraphics[height=4.0cm]{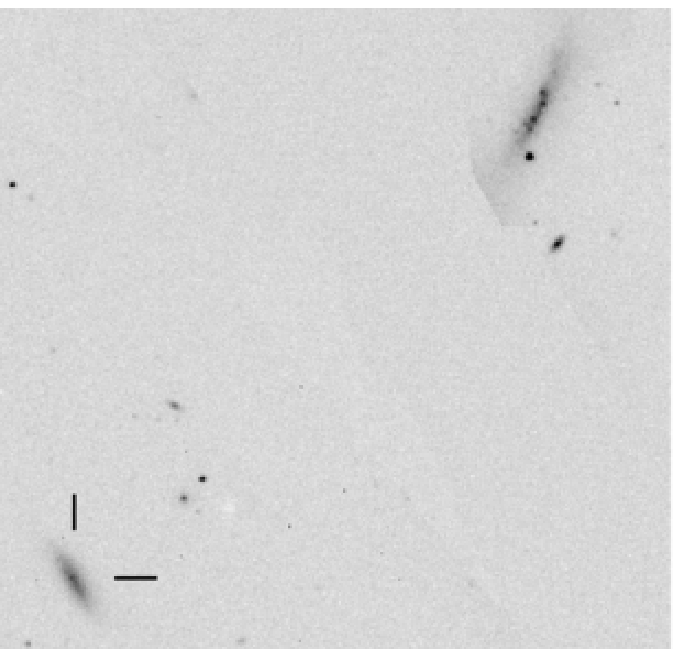}

\vspace*{0.4cm}
\includegraphics[height=4.0cm]{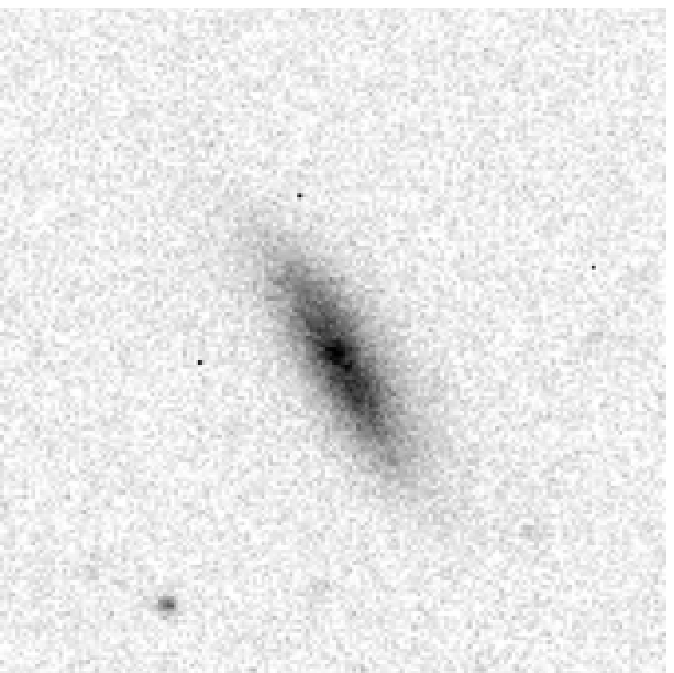}
\includegraphics[height=4.0cm]{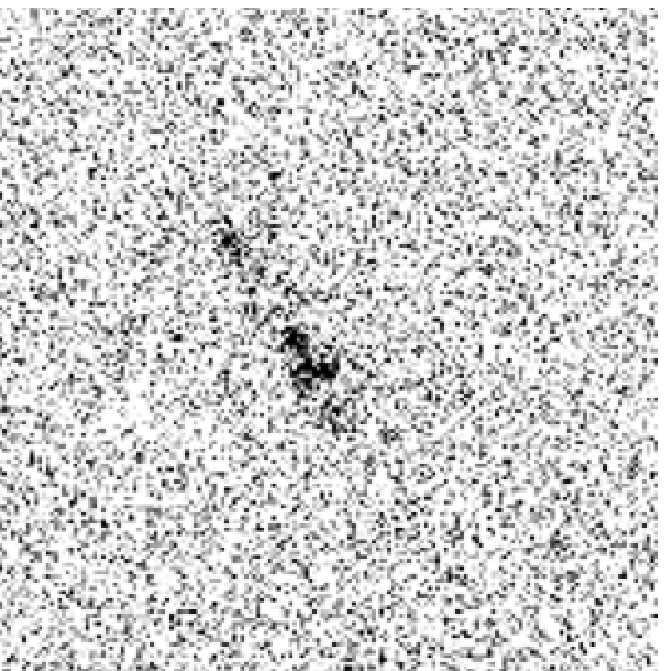}
\end{center}
\caption{
Upper image: UGC~6506 and satellite galaxy (image size 212$^{\prime \prime}$ by
206$^{\prime \prime}$).  The central part of this image
suffered contamination from a band of scattered light in a diagonal
strip between the two galaxies; this has been removed in the data
reduction, along with some stars which lie in the centre of this
field.  The relative brightness and orientation of the two galaxies
are not affected.
Lower images: Satellite galaxy in $R$-band (left) and continuum-subtracted 
\Ha\ (right: image sizes 54$^{\prime \prime}$ by 54$^{\prime \prime}$).
}
\label{fig:u6506}
\end{figure}

\vspace*{1.0 cm}

\begin{figure}
\begin{center}
\includegraphics[height=4.0cm]{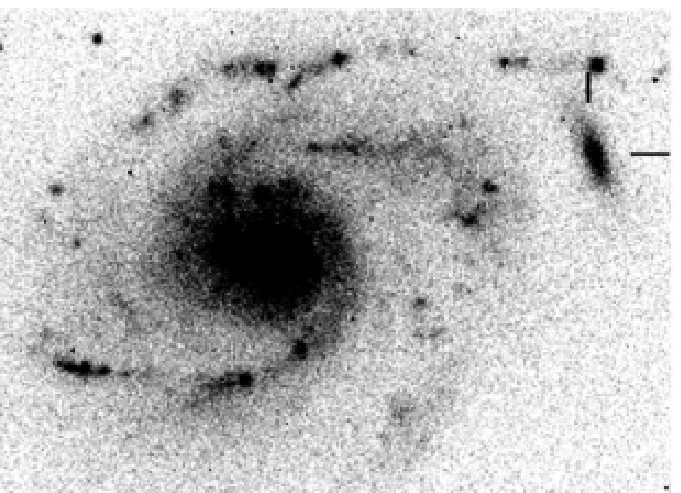}

\vspace*{0.4cm}
\includegraphics[height=4.0cm]{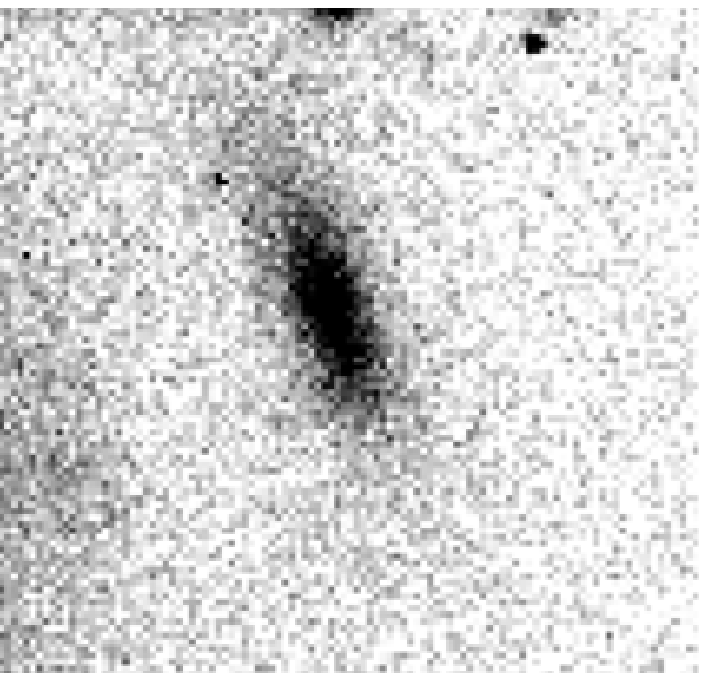}
\includegraphics[height=4.0cm]{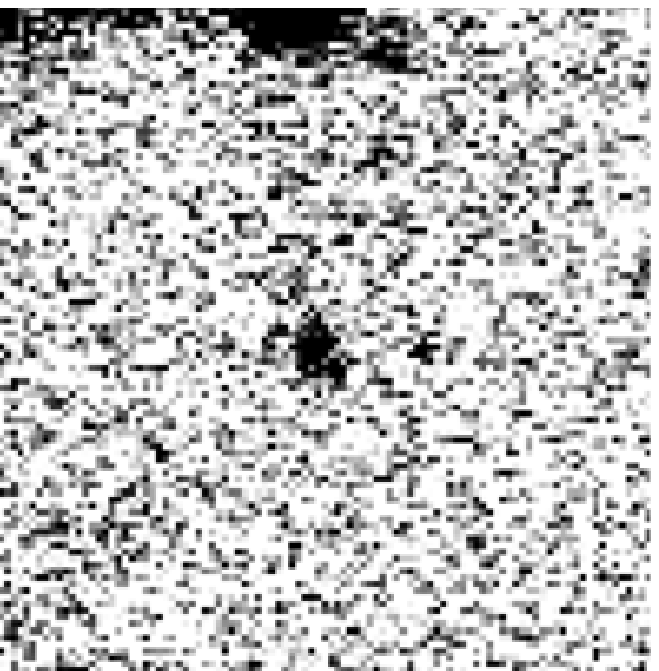}
\end{center}
\caption{
Upper image: UGC~12788 and satellite galaxy (image size 113$^{\prime
\prime}$ by 82$^{\prime \prime}$).  
Lower images: Satellite galaxy in $R$-band (left) and continuum-subtracted 
\Ha\ (right: image sizes 33$^{\prime \prime}$ by 33$^{\prime \prime}$).
}
\label{fig:u12788}
\end{figure}

\end{appendix}

\end{document}